\documentclass[fleqn,usenatbib]{mnras}
\usepackage{newtxtext,newtxmath}
\usepackage[T1]{fontenc}
\usepackage{ae,aecompl}
\usepackage{mathtools}
\usepackage{asymptote}
\usepackage{comment}

\usepackage{graphicx}	% Including figure files
\usepackage{amsmath}	% Advanced maths commands
\def\spa{s_\parallel}
\def\spe{s_\perp}
\def\der{\mathrm{d}}

\def\apar{\alpha_\parallel}
\def\aper{\alpha_\perp}

\title[Clustering with nulled angular modes]{Angular systematics-free cosmological analysis of galaxy clustering in configuration space}

\author[Paviot et al.]{\parbox{\textwidth}{
Romain Paviot$^{1,2}$\thanks{E-mail: romainpaviot@gmail.com},
Sylvain de~la~Torre$^{1}$,
Arnaud de~Mattia$^{3}$,
Cheng Zhao$^{4}$,
Julian Bautista$^{2}$,
Etienne Burtin$^{3}$,
Kyle Dawson$^{5}$,
St\'ephanie Escoffier$^{2}$,
Eric Jullo$^{1}$,
Anand Raichoor$^{4}$,
Ashley J. Ross$^{6}$,
Graziano Rossi$^{7}$
}
\vspace*{10pt} \\
$^{1}$ Aix Marseille Univ, CNRS, CNES, LAM, Marseille, France \\
$^{2}$ Aix Marseille Univ, CNRS/IN2P3, CPPM, Marseille, France \\
$^{3}$ IRFU, CEA, Universit\'e Paris-Saclay, F-91191 Gif-sur-Yvette, France \\
$^{4}$ Institute of Physics, Laboratory of Astrophysics, \'Ecole Polytechnique F\'ed\'erale de Lausanne (EPFL), Observatoire de Sauverny, 1290 Versoix, Switzerland \\
$^{5}$ Department of Physics and Astronomy, 
University of Utah, Salt Lake City, UT 84112, USA \\
$^{6}$ Center for Cosmology and Astro-Particle Physics
Ohio State University, Columbus, OH 43210, USA \\
$^{7}$ Department of Physics and Astronomy, 
Sejong University, Seoul, 143-747, Korea
}

\pubyear{2021}

\begin{document}
\label{firstpage}
\pagerange{\pageref{firstpage}--\pageref{lastpage}}
\maketitle

\begin{abstract}
Galaxy redshift surveys are subject to incompleteness and inhomogeneous sampling due to the various constraints inherent to spectroscopic observations. This can introduce systematic errors on the summary statistics of interest, which need to be mitigated in cosmological analysis to achieve high accuracy. Standard practices involve applying weighting schemes based on completeness estimates across the survey footprint, possibly supplemented with additional weighting schemes accounting for density-dependent effects. In this work, we concentrate on pure angular systematics and describe an alternative approach consisting in analysing the galaxy two-point correlation function where angular modes are nulled. By construction, this procedure removes all possible known and unknown sources of angular observational systematics, but also part of the cosmological signal. We use a modified Landy-Szalay estimator for the two-point correlation function that relies on an additional random catalogue where angular positions are randomly drawn from the galaxy catalogue, and provide an analytical model to describe this modified statistic. We test the model by performing an analysis of the full anisotropic clustering in mock catalogues of luminous red and emission-line galaxies at $0.43<z<1.1$. We find that the model fully accounts for the modified correlation function in redshift space, without introducing new nuisance parameters. The derived cosmological parameters from the analysis of baryon acoustic oscillations and redshift-space distortions display slightly larger statistical uncertainties, mostly for the growth rate of structure parameter $f\sigma_8$ that exhibits a $50\%$ statistical error increase, but free from angular systematic error. 
\end{abstract}

\begin{keywords}
galaxies: statistics -- large-scale structure of Universe.
\end{keywords}

%%%%%%%%%%%%%%%%%%%%%%%%%%%%%%%%%%%%%%%%%%%%%%%%%%

%%%%%%%%%%%%%%%%% BODY OF PAPER %%%%%%%%%%%%%%%%%%

\section{Introduction}

The mapping of the large-scale structure with galaxy spectroscopic surveys is a major source of information for determining the cosmological model. This owes to its unique sensitivity to both the background expansion and growth of density perturbations through gravity \citep[e.g.][]{alam21}. Galaxy spectroscopic surveys have played a crucial role in the last two decades in providing increasingly precise measurements of the cosmological parameters  \citep[e.g.][]{peacock01, cole05,tegmark06,percival10,blake12,pezzotta17,alam17,alam21}. This has been possible particularly thanks to the development of multi-object spectrographs, whose improved 
multiplexing capabilities and use in large programs have allowed a tremendous increase of the surveyed cosmological volume \citep[e.g.][]{lewis02,lefevre03,gunn06,Smee2013}. Nonetheless, extracting genuine spatial information from those surveys can be challenging because of the complexity of the selection function. This is particularly crucial for cosmological inference, where a significant effort is necessary to control all observational systematic effects and avoid degrading the accuracy on the derived cosmological parameters \citep[e.g.][]{blake10,ross12,delatorre13,reid16,ross20}.

Galaxy spectroscopic surveys are generally exposed to incompleteness for various reasons. In multi-slit or multi-fibre spectroscopic surveys, incompleteness can be induced by missing observations resulting from the mechanical limitations of the spectrograph. For instance, the finite size of fibres (or slits) and their finite usable number at each observation prevent spectroscopic observations of all possible targets. The latter aspect also depends on the observational strategy, particularly the number density of targets and redundancy at observing the same patches of the sky. This incompleteness can be strongly correlated with the intrinsic clustering of targets, as in the case of fibre collision for instance. Another source of incompleteness includes varying foreground or background noises such as stellar density or galactic extinction, which aggravate our ability to extract redshift measurements from observed spectra, or lead to very low signal-to-noise spectra where no redshift can be determined. Inhomogeneity in the survey sampling can also arise in the preparation of the target sample, when for instance targeted sources are not selected in the same way in different patches of the sky due to limited or uneven photometry. Overall, these effects result in systematic biases on clustering measurements, which in turn, can introduce biases in the inferred values of cosmological or physical parameters.

In general, observational systematic errors are not necessarily all known but need to be mitigated in cosmological analysis in order to achieve high accuracy. Standard mitigation strategies involve applying weighting schemes based on completeness estimates across the survey footprint and additional schemes to account for projected density-dependent effects. In the Baryon Oscillation Spectroscopic Survey \citep[BOSS,][]{boss} and extended Baryon Oscillation Spectroscopic Survey \citep[eBOSS,][]{eboss} for instance, completeness weights were calculated by tessellating the observed sky and fitting multilinear regression to residual trends in observational parameters, such as star density or survey depth \citep{ross12}. In addition, projected density-dependent effects such as fibre collision were dealt with by up-weighting nearest neighbour galaxies to missed galaxy. In this work, we describe an alternative approach that consists in analysing the galaxy two-point statistics in configuration space, with nulled angular modes. This is possible by modifying the standard estimator of the two-point correlation function. This idea was first introduced by \citet{burden17}. They proposed an estimator similar to the standard \citet{landy93} estimator, but that includes an additional random catalogue where angular positions are randomly drawn from the galaxy catalogue. This effectively permits removing the angular clustering, as angular correlations are canceled by the new random catalogue. The amplitude of this new statistic is suppressed with respect to the standard two-point correlation function, but is blind to any systematic angular selection effects. A similar method was also developed in Fourier space in \citet{pinol17}. The first application to real data of such estimator was performed on the eBOSS emission-line galaxy sample by \citet{tamone20}. 

In this work, we derive a full model for the modified two-point correlation function in redshift space, assess its accuracy, and perform a full analysis of baryon acoustic oscillations (BAO) and redshift-space distortions (RSD) on luminous red galaxy and emission-line galaxy mock samples, as a proof of concept. We will refer to this modified statistic as the angular modes-free (AMF) two-point correlation function. 

The paper is organised as follows. The Section 2 presents the formalism of the AMF two-point correlation function. The corresponding theoretical model is presented and tested against mock galaxy samples in Section 3. BAO and RSD analyses are performed using the standard and AMF correlation functions in Section 4. Section 5 discusses the results and conclude.

%Throughout this work we adopt the $\Lambda$CDM fiducial cosmology used in eBOSS DR16 cosmological analyses \citep{alam21}, i.e. $\Omega_m=0.31$, $\Omega_\Lambda=0.69$, $\Omega_bh^2=0.022$, $h=0.676$, $\sigma_8=0.8$ and $n_s=0.97$.

\section{The angular modes-free correlation function}

\subsection{Definition}

The cosmological information in galaxy redshift surveys is commonly extracted from the measured two-point statistics of the galaxy spatial distribution. In configuration space, this is achieved using the minimum variance \citet{landy93} estimator:
\begin{equation} \label{eq:LS}
\xi(\mathbf{s}) = \frac{DD(\mathbf{s}) - 2DR(\mathbf{s}) + RR(\mathbf{s})}{RR(\mathbf{s})},
\end{equation}
where $\mathbf{s}$ is the separation vector between pairs of objects and $DD$, $DR$ and $RR$ are respectively the normalised number of galaxy-galaxy, galaxy-random, and random-random pairs. The random catalogue consists in random points uniformly distributed over the survey footprint and with the same radial distribution as the data. The separation vector can be decomposed into $(s,\mu)$ coordinates, where $s$ is the norm of the separation vector $\mathbf{s}$ and $\mu$ is the cosine angle between the separation and line-of-sight directions. This decomposition enables the expansion of the correlation function in multipole moments,
\begin{equation} \label{eq:xil}
\xi_\ell(s) = \frac{(2\ell+1)}{2} \int_{-1}^{1} \xi(s,\mu) L_\ell(\mu)\rm{d} \mu,
\end{equation}
where $L_\ell$ is the Legendre polynomial of order $\ell$. 

In order to suppress the angular modes, one can modify this estimator by introducing an auxiliary random catalogue. The latter has exactly the same angular clustering pattern as the data but a random realisation of the radial distribution. It is easily constructed by randomly assigning galaxy angular positions from the data catalogue to random points. We will refer to it as the shuffled random catalogue, $S$, in the following. One can thus design a modified \citet{landy93} estimator such that \citep{burden17}:
\begin{equation} \label{eq:LSM}
\tilde{\xi}(\mathbf{s}) = \frac{DD(\mathbf{s}) - 2DS(\mathbf{s}) + SS(\mathbf{s})}{RR(\mathbf{s})}.
\end{equation}
In this estimator the standard random catalogue in the numerator is replaced by the shuffled random catalogue, while is kept in the denominator. By imprinting the angular pattern of the galaxies in the random catalogue $S$, one suppresses the angular clustering and associated potential systematic errors, but at the price of removing part of the cosmological information. Similarly to the standard correlation function, the AMF correlation function can be expanded in multipole moments.

The purpose of the random catalogues is the estimation of the observed volume, and for this it must contain a large number of points (typically 20-50 times more than in the galaxy catalogue). In the case of the shuffled random catalogue, since angular positions are drawn from observed galaxy positions, some angular positions will be repeated. This leads to some $SS$ or $DS$ pairs with vanishing angular separation, or equivalently with $\mu=1$. Keeping these pairs in the pair counts can introduce additional noise and bias in the estimation of the AMF correlation function multipole moments. In practice however, by adopting a proper binning in $\mu$ in the pair counts, these pairs can be discarded. This effect is more problematic in Fourier space where $\mu=1$ associated modes cannot be discarded in the estimator and introduce an additional shot-noise term, 
as discussed in \citet{demattia19}.

\subsection{Modelling}

In order to model the AMF correlation function, we follow \citet{burden17} and define the AMF overdensity field
\begin{equation}
\tilde{\delta}(\mathbf{r}) \equiv \frac{n(\mathbf{r}) - \tilde{n}(\mathbf{r})}{\bar{n}(\mathbf{r})}
\end{equation}
where $n(\mathbf{r})$, $\tilde{n}(\mathbf{r})$ and ${\bar{n}(\mathbf{r})}$ correspond respectively to the number density of galaxies, shuffled random points, and standard random points at comoving position $\mathbf{r}$. By construction, the shuffled random number density is 
\begin{equation}
\tilde{n}(\mathbf{r}) = \frac{\int n(\mathbf{\chi',\gamma}) \der \chi' \int{\bar{n}(\mathbf{\chi,\gamma'}) \der \gamma'}}{\int \int \bar{n}(\mathbf{\chi',\gamma'}) \der \chi' \der \gamma'} = 
\bar{n}(\mathbf{r}) \frac{\int n(\mathbf{\chi',\gamma}) \der \chi'}{\int \bar{n}(\mathbf{\chi',\gamma'}) \der \chi'},
\end{equation}
where $\gamma$ corresponds to the two-dimensional angular coordinates and $\chi$ to the radial coordinate. We assumed that the random catalogue is uniform across the sky in $\gamma$. We can thus express the AMF overdensity field as
\begin{equation} \label{eq:deltashuf}
\tilde{\delta}(\mathbf{r}) = \delta(\mathbf{r}) - \frac{\int {\delta(\chi,\gamma) \bar{n}(\chi) \der \chi}}{\int \bar{n}(\chi) \der \chi}.
\end{equation}
In the following, in order to simplify the notation, $\bar{n}$ is normalized so that $\int \bar{n}(\chi) d\chi=1$. In Eq. \ref{eq:deltashuf}, the second term on the right-hand side corresponds in fact to the projected overdensity at angular position $\gamma$ on the sky,
\begin{equation}
    \hat{\delta}(\gamma)= \int {\delta(\chi,\gamma) \bar{n}(\chi) \der \chi}.
\end{equation}
The AMF correlation function corresponds to the auto-correlation of the AMF overdensity field,
\begin{multline} \label{eq:shuf}
\tilde{\xi}(\mathbf{s}) \equiv \langle \tilde{\delta}(\mathbf{r}) \tilde{\delta}(\mathbf{r'}) \rangle
= \langle \delta(\mathbf{r}) \delta(\mathbf{r'}) \rangle - 2\left\langle 
\delta(\mathbf{r}) \hat{\delta}(\gamma') \right\rangle + \left\langle 
\hat{\delta}(\gamma) \hat{\delta}(\gamma') \right\rangle,
\end{multline}
where we have defined $\mathbf{s} = \mathbf{r'}-\mathbf{r}$ and $\langle . \rangle$ denotes the ensemble average. In the latter equation, the first and third terms correspond respectively to the three-dimensional and angular correlation functions, while the second term is the cross-correlation between the three-dimensional overdensity and projected angular overdensity fields.

The angular correlation term is defined as
\begin{equation}
w(\theta) \equiv \left \langle \int {\delta(\chi,\gamma) \bar{n}(\chi) \der \chi} \int {\delta(\chi',\gamma') \bar{n}(\chi') \der \chi'} \right \rangle
\end{equation}
and is related to the three-dimensional correlation function $\xi$ as \citep{peebles80},
\begin{equation}
    w(\theta) = \int_0^\infty \der \chi \int_0^\infty \der \chi' \xi(\theta,\Delta \chi)  \bar{n}(\chi) \bar{n}(\chi'),
\end{equation}
where $\theta=|\gamma'-\gamma|$ is the angular separation and $\Delta \chi=\chi'-\chi$ is the radial separation. Further defining $\bar{\chi}=(\chi+\chi')/2$ and changing the variable of integration in the integrals leads to \citep{simon07}
\begin{equation} \label{eq:wthetatrue}
    w(\theta) = \int_0^\infty \der \bar{\chi} \int_{-2\bar{\chi}}^{2\bar{\chi}} \der \Delta \chi \, \xi(\theta,\Delta \chi)  \bar{n}\left(\bar{\chi}-\frac{\Delta \chi}{2}\right) \bar{n}\left(\bar{\chi}+\frac{\Delta \chi}{2}\right).
\end{equation}
If we assume that $\bar{n}$ weakly varies over the typical $\Delta \chi/2$ scale in $\xi(\theta,\Delta \chi)$ so that $\bar{n}\left(\bar{\chi}-\Delta \chi/2\right) \simeq \bar{n}\left(\bar{\chi}+\Delta \chi/2\right) \simeq \bar{n}(\chi)$ in the inner integral, Eq. \ref{eq:wthetatrue} simplifies to the well-known Limber approximation \citep{limber53}:
\begin{equation} \label{eq:angtermapprox}
w(\theta) \simeq  \int_0^{\infty} \der \bar{\chi} \, \bar{n}^{2}(\chi) \int_{-\infty}^{\infty} \der \Delta \chi \, \xi(\theta,\Delta \chi).
\end{equation}

The cross-correlation term can be written as \citep{burden17}
\begin{equation} \label{eq:crossterm}
\left\langle \delta(\chi,\gamma) \int {\delta(\chi',\gamma') \bar{n}(\chi') d\chi'}
\right\rangle \simeq \int_0^\infty \der \chi' \, \bar{n}(\chi') \, \xi(\theta,\Delta \chi),
\end{equation}
and by adopting the same changes of variable as previously, the right-hand side of Eq. \ref{eq:crossterm} becomes
\begin{equation} \label{eq:crosstermapprox}
\int_{-\infty}^\infty \der \Delta \chi \, \bar{n}\left(\bar{\chi} + \frac{\Delta \chi}{2}\right) \, \xi(\theta,\Delta \chi).
\end{equation}
In this equation, $\bar{\chi}$ is an undefined constant, which is related at first order to the mean radial distance of the sample. The right-hand side of Eq. \ref{eq:crossterm} used to obtain this result should in reality be averaged over the observed volume, introducing a further integral over $\der^3 \chi$. In fact, this approximation can be avoided by making explicit the volume integral over the survey window function as shown in Section \ref{sec:fullmodel}. We note that if we use Limber-type approximation in Eq. \ref{eq:crosstermapprox}, the mean number density exit the integral and the expression reduces to a constant times the projected correlation function.

In practice, we are seeking an expression for the anisotropic three-dimensional correlation function that can be used to model observed multipole moments. In the plane-parallel approximation, the separation vector $\mathbf{s}$ can be decomposed in terms of the transverse and radial comoving separations, $\spe$ and $\spa$ respectively, using for instance the mid-point line-of-sight definition \citep{fisher94}. We can thus use the previous model defined for $\xi(\theta,\Delta \chi)$ and make the substitutions: $\Delta \chi \rightarrow \spa$ and $\theta \rightarrow \spe$. This holds when the radial distance is large with respect to the pair separation and effectively assumes a flat sky. In this case, we obtain that
\begin{align} \label{eq:modelflatsky}
    \tilde{\xi}(\spe,\spa) &= \xi(\spe,\spa) - 2 C(\spe) + A(\spe)
\end{align}
where
\begin{align}
C(\spe) &=  \int_{-\infty}^\infty \der \spa \,  \bar{n}\left(\bar{\chi}+\frac{\spa}{2}\right) \, \xi(\spe,\spa), \label{eq:fmodelC} \\
A(\spe) &= \int_0^{\infty} \der \bar{\chi} \int_{-\infty}^{\infty} \der \spa \xi(\spe,\spa) \, 
\bar{n}\left(\bar{\chi}-\frac{\spa}{2}\right) \bar{n}\left(\bar{\chi}+\frac{\spa}{2}\right),
\label{eq:fmodelA}
\end{align}
or with Limber approximation,
\begin{align}
C_{\rm L}(\spe) &= \bar{n}(\bar{\chi}) \, \int_{-\infty}^\infty \der \spa \,  \xi(\spe,\spa), \label{eq:fmodelCL} \\
A_{\rm L}(\spe) &= \int_0^{\infty} \der \bar{\chi} \bar{n}^{2}(\bar{\chi}) \int_{-\infty}^{\infty} \der \spa \xi(\spe,\spa). \label{eq:fmodelAL}
\end{align}
By substituting $A$ and $C$ by $A_{\rm L}$ and $C_{\rm L}$ in Eq. \ref{eq:modelflatsky} one defines the simplest model, where both Limber and flat-sky approximations are used. We note that in the analysis of \citet{tamone20}, such model is used where $C$ and $A_{\rm L}$ are taken as cross-correlation and angular terms, respectively. In those approximate models, $\bar{\chi}$ is a free parameter that can be determined empirically from simulations for the specific galaxy sample under consideration. Eventually, the AMF correlation function multipole moments can be obtained by remapping $\tilde{\xi}(\spe,\spa)$ into $\tilde{\xi}(s,\mu)$ using that $\spe = s \sqrt{1-\mu^2}$ and $\spa = s \mu$, and integrating $\tilde{\xi}(s,\mu)$ over $\mu$ as in Eq. \ref{eq:xil}.

\subsection{Full model} \label{sec:fullmodel}

In fact, the flat-sky approximation and that made in Eq. \ref{eq:crossterm} can be avoided. Precisely, the AMF estimator in Eq. \ref{eq:LSM} corresponds to the auto-correlation of the AMF overdensity times the survey window function, divided by the survey window correlation function. The survey window function $P(\mathbf{r})$ is the probability of seeing an object at any position $\mathbf{r}$ in the survey. If we define the windowed AMF overdensity field $F(\mathbf{r})$ as
\begin{equation}
    F(\mathbf{r}) = P(\mathbf{r})\delta(\mathbf{r}) - P(\mathbf{r}) \int \der r' \bar{n}(r') \delta(\mathbf{r}'), 
\end{equation}
where $\mathbf{r}$ and $\mathbf{r}'$ are collinear, the AMF correlation function is
\begin{equation}
    \tilde{\xi}(\mathbf{s}) \equiv \frac{\int \der^3 r \, F(\mathbf{r}) F(\mathbf{r}+\mathbf{s})}{\int \der^3 r \, P(\mathbf{r}) P(\mathbf{r}+\mathbf{s})}.
\end{equation}
The expected value of the latter estimator is
\begin{equation} \label{eq:fullmodel}
    \tilde{\xi}(\mathbf{s}) = \xi(\mathbf{s}) - \frac{C(\mathbf{s})}{W(\mathbf{s})} +
    \frac{A(\mathbf{s})}{W(\mathbf{s})},
\end{equation}
where $\xi(\mathbf{s})$ is the standard anisotropic correlation function and 
\begin{align}
    C(\mathbf{s}) &= \int \der^3 r \, P(\mathbf{r}) P(\mathbf{r}+\mathbf{s}) \left[ 
    \int \der r'  \bar{n}(r') \xi(\mathbf{r}'-\mathbf{r}) \right. \\
    \nonumber
   &+ \left. \int \der r''  \bar{n}(r'') \xi\left(\mathbf{r}''-\mathbf{r}-\mathbf{s}\right) \right], \\
    A(\mathbf{s}) &= \int \der^3 r \, P(\mathbf{r}) P(\mathbf{r}+\mathbf{s})
    \int \der r''  \bar{n}(r'')
    \int \der r'  \bar{n}(r') \xi(\mathbf{r'}-\mathbf{r''}), \\
    W(\mathbf{s}) &= \int \der^3 r \, P(\mathbf{r}) P(\mathbf{r}+\mathbf{s}).
\end{align}

The geometrical configuration and details of the derivation are given in Appendix A, we only summarise here the main results. The integrals in the expressions for $A(\mathbf{s})$ and $C(\mathbf{s})$ can be simplified and rearranged, and in the end we find that the AMF correlation function reads
\begin{equation}
\label{eq:shuf_model}
    \tilde{\xi}(s,\mu) = \xi(s,\mu) - \frac{C(s,\mu)}{W(s,\mu)} +
    \frac{A(s,\mu)}{W(s,\mu)}
\end{equation}
where
\begin{align}
    C(s,\mu) &= \sum_{\ell=0}^{\infty} \left( \int_0^{\infty} \der\Delta \sum_{p=0}^{\infty} \Delta \, \xi_p(\Delta) \mathcal{W}_{C \ell p}(s,\Delta)\right) L_\ell(\mu) \label{eq:Csmufmodel}\\
    A(s,\mu) &= \sum_{\ell=0}^{\infty} \left( \int_0^{\infty} \der\Delta \sum_{p=0}^{\infty} \Delta \, \xi_p(\Delta) \mathcal{W}_{A \ell p}(s,\Delta) \right) L_\ell(\mu) \label{eq:Asmufmodel}\\
    W(s,\mu) &= \sum_{\ell=0}^{\infty} \mathcal{W}_{\ell}(s) L_\ell(\mu).
\end{align}
We refer to Appendix A for the expression of the kernels $\mathcal{W}_{C \ell p}$, $\mathcal{W}_{A \ell p}$, and $\mathcal{W}_{\ell}$. It is important to emphasize that these kernels only depend on the geometry of the survey and can be computed independently. Eventually, the multipole moments of the AMF correlation function are obtained from $\tilde{\xi}(s,\mu)$ as
\begin{equation} \label{eq:multipolefmodel}
\tilde{\xi}_\ell(s) = \frac{(2\ell+1)}{2} \int_{-1}^{1} \tilde{\xi}(s,\mu) L_\ell(\mu) \der \mu.
\end{equation}
Contrary to the approximate models presented in the previous section, the full model does not include any approximation except the plane-parallel one, and has no additional free parameter.

\subsection{Sensitivity to angular systematics}

\begin{figure} 
     \centering
     \includegraphics[width=1.0\columnwidth]{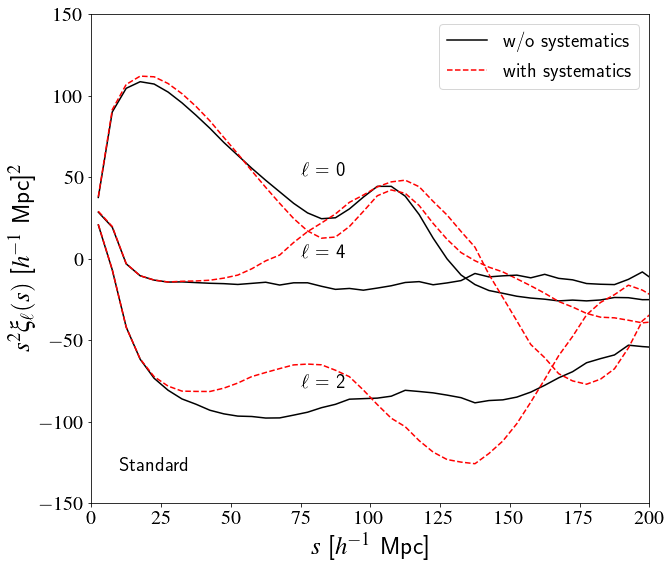}
     \includegraphics[width=1.0\columnwidth]{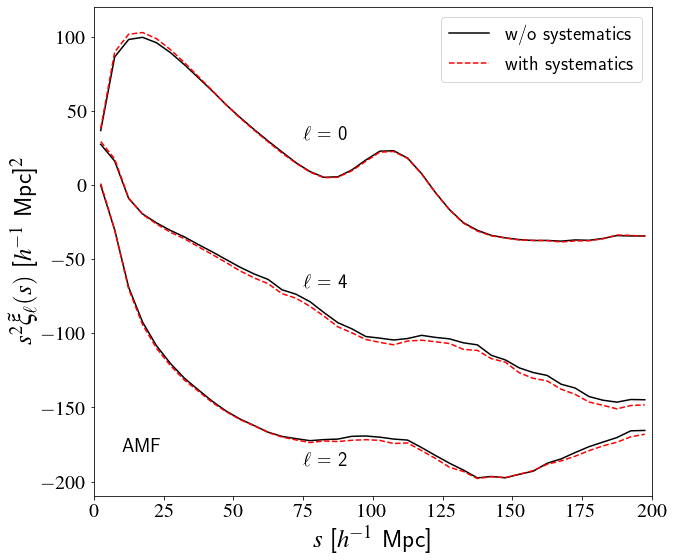}
     \caption{CMASS mocks mean measurements of the standard (top figure) and AMF (bottom figure) correlation function multipole moments with (solid lines) and without (dashed lines) angular systematics.}
   \label{fig:2CPFwithsyst}  
\end{figure}

It is interesting to see formally that the AMF overdensity removes any additive angular contamination. Indeed, if we write the contaminated overdensity $\delta(\mathbf{r})+c(\mathbf{r})$, where $c$ is a contamination field that only depends on the line-of-sight direction, we have for the windowed AMF overdensity,
\begin{align}
    F(\mathbf{r}) 
            & = P(\mathbf{r})\left(\delta(\mathbf{r})+c(\mathbf{r})\right) - P(\mathbf{r}) \int \der r' \bar{n}(r') \left(\delta(\mathbf{r}')+c(\mathbf{r}')\right) \\
            &= P(\mathbf{r})\delta(\mathbf{r}) - P(\mathbf{r}) \int \der r' \bar{n}(r') \delta(\mathbf{r}'). 
\end{align}
To obtain the latter equation we have used that, by definition of $c$ and the fact that $\mathbf{r}$ and $\mathbf{r}'$ share the same line of sight, $c(\mathbf{r}')=c(\mathbf{r})$. Nonetheless, if the contamination field modulates the observed number of galaxies, as for instance in the case of varying survey depth or galactic extinction \citep[][]{shafer15}, both additive and multiplicative components will arise, and the multiplicative one will not be erased (it will factorise $F(\mathbf{r})$).

In order to verify previous statements, we test the efficiency of the AMF two-point correlation function estimator at removing a spurious angular modulation in galaxy number using the galaxy CMASS mocks introduced in Section \ref{sec:mocks}. We instil an artificial angular modulation in the number of galaxies $N$, such that $N \, \rightarrow \, N \left(1+\epsilon\right)$, in a similar fashion as in \cite{demattia19}. This is done by weighting each galaxy at angular position (RA, Dec) by 
\begin{equation}
\label{eq:weightsangular}
w_{\text{ang}}(\text{RA}, \text{Dec})= 1 + 0.2\sin\left(\frac{2\pi}{10}\times\text{RA}\right) \sin\left(\frac{2\pi}{5}\times \text{Dec}\right).
\end{equation}
We choose here a modulation amplitude of $20\%$, which is typical of the observed level. The standard and AMF correlation function multipole moments are measured in each mock and later averaged, including or not galaxy weights $w_{\text{ang}}$. In the standard two-point correlation function estimator, the angular weights are only applied to the galaxy catalogue, while in the AMF estimator, they are applied in the galaxy and shuffled random catalogues. The resulting measurements are shown in Fig. \ref{fig:2CPFwithsyst}. We can see for the standard two-point correlation function that angular systematics significantly affect all the multipole moments. On the other hand, the AMF multipole moments are nearly unaffected by angular systematics. We only see a negligible shift in amplitude on the small scales of the monopole and on the hexadecapole. This demonstrates that the multiplicative component of the angular contamination is very small, and most of the effect of the angular modulation is removed in the AMF estimator.

%These shifts simply correspond to a wrong normalization, as given by the standard random catalog, which no longer properly describe the survey window function, when angular weights are introduced in the mocks. Thus, an approximate treatment of the angular systematic should be enough to give unbiased measurements of the AMF 2PCF.

\section{Test on mock samples} \label{sec:mocks}

\begin{figure} 
     \centering
     \includegraphics[width=\columnwidth]{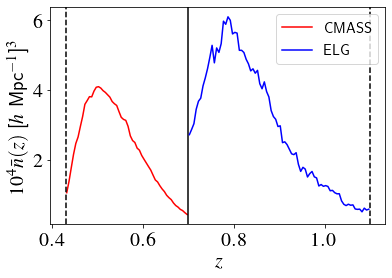}
     \caption{Galaxy number density in CMASS (red curve) and ELG (blue curve) mock samples.}
   \label{fig:nz_sample}
\end{figure}

\subsection{Description of the mock samples}

In order to test the AMF full model presented in the previous section, we make use of NSeries mocks \citep{alam17} and eBOSS EZmocks \citep{zhao20}. These mocks are designed to reproduce different galaxy samples of the BOSS and eBOSS surveys. We concentrate in this analysis on the BOSS constant mass galaxy (CMASS) and eBOSS emission-line galaxy (ELG) samples, which cover the redshift intervals $0.43 < z < 0.7$ and $0.7<z<1.1$ respectively.

The NSeries galaxy mocks are based on a N-body simulation populated with a single halo occupation distribution (HOD) model. These were built to reproduce the observed North
Galactic Cap (NGC) subset of the BOSS CMASS galaxy sample, which covers the redshift range $0.43 < z < 0.7$  and has an effective redshift of $z_{\rm eff} = 0.55$. There are 84 mocks in total, which have a very realistic small-scale clustering imprinted.
We refer the reader to \citet{alam17} for the detailed description of these
mocks. Given the modest number of mock realisations at disposal, we do not use them directly to estimate correlation function covariance matrices. Instead, the latter are estimated from $2048$ realisations of the same volume based on \textsc{Patchy} approximated method \citep[][]{kitaura14}.

%The BOSS CMASS NGC mocks cover 7000 $\deg^2$ for a total effective volume of approximately 2.4 Gpc$^3$, for each individual realization. 

%The Ezmocks mocks 
The EZmocks were built by gravitationally evolving dark matter particles with Zel\'\ dovich approximation. They include a nonlinear, non-local, and scale-dependent galaxy bias prescription allowing the addition of galaxies on top of the dark matter field. The redshift-space two-point statistics in these mocks agrees with N-body simulation within $1\%$ down to $10~h^{-1}~{\rm Mpc}$ \citep{Chuang2015}. We use in this analysis $500$ ELG EZmocks that mimic the geometry and observed clustering of the eBOSS ELG South Galactic Cap (SGC) dataset. These mocks cover the redshift range  $0.7<z<1.1$ and have an effective redshift of $z_{\rm eff} = 0.86$. We refer the reader to \citet{zhao20} for the detailed description of these mocks.

We measure in all the mocks the standard and AMF correlation functions in redshift space using the estimators in Eqs. \ref{eq:LS} and \ref{eq:LSM}, respectively. We used random catalogues with approximately $50$ times the number of 
galaxies in the mock data. The shuffled random catalogues have randomly drawn angular positions from the mock data catalogues, but the same radial distribution as that imprinted in the standard random catalogues.

%, hence, we do not include the radial integral constraint correction \citep[][]{demattia19}.

\begin{table}
    \centering
    \caption{Cosmological parameters of the {\sc EZmocks} (EZ) and the {\sc Nseries}(NS) simulation. For both simulation $\Omega_{\nu}= 0$. The effective redshift of the {\sc EZmocks} and {\sc Nseries} are $z_{\text{eff}}=0.86$ and $z_{\text{eff}}=0.55$ respectively.
    }
    \label{tab:cosmologies_simu}
    \begin{tabular}{ccc}
    \hline
    \hline
    &  \sc EZ  & \sc NS \\
    \hline
$\Omega_m$& 0.307 & 0.286   \\
$\Omega_b$& 0.048 & 0.047   \\
$h$& 0.678 & 0.700  \\
$n_s$& 0.961 & 0.960  \\
$\sigma_8(z=0)$ & 0.823 & 0.820 \\
$r_{\rm drag}$ [Mpc] & 147.66 & 147.15 \\
$f\sigma_8(z=z_{\text{eff}})$ & 0.469 & 0.449 \\
    \hline
    \end{tabular}
\end{table}

\subsection{Implementation of the full model}

We use a similar method as that presented in \citet{breton21} to calculate the full model kernels. We first build angular {\sc healpix} \citep[][]{Gorski2005} maps from the survey footprints used to create the mocks. 
These maps are used to estimate the angular selection correlation function, $\Phi(\theta)$, with {\sc polspice} code
\citep[][]{Szapudi2001,Chon2004}. The $\bar{n}(\chi)$ are estimated from
CMASS and ELG random catalogues and shown in Fig. \ref{fig:nz_sample}. Redshifts are converted to comoving distances using the corresponding fiducial cosmology of the simulation, given in Table \ref{tab:cosmologies_simu}. From these two ingredients, the kernels can be evaluated numerically using multi-dimensional Monte-Carlo integration methods. Specifically, we use the {\sc cuba} library \citep[][]{hahn04} in a similar way as in \citet{breton21} to solve numerically the kernel integrals given in Appendix A. A code to compute those kernels for any survey geometry is publicly available\footnote{\url{https://github.com/mianbreton/RR_code}} Once kernels are computed, the cross-correlation and angular terms are obtained by integrating over $\Delta$ the kernels times the model standard correlation function. In practice, this integral is performed as a Riemmann sum. We find that a $\Delta$ binning of $1\;$Mpc/h is sufficient to have a numerically stable model estimation. 

\subsection{Test of the models}

In order to assess the different models presented previously, we compare their prediction to the mean AMF correlation function measured in the CMASS mocks. The models take as input the redshift-space galaxy correlation function and number density as a function of radial distance. For the purpose of testing AMF models, we fix those to their mean mocks values. 

\begin{figure} 
     \centering
     \includegraphics[width=\columnwidth]{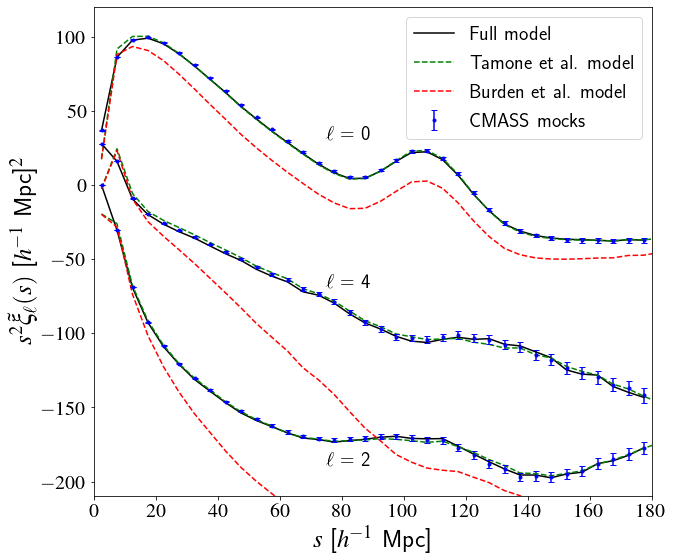}
     \caption{Comparison of AMF correlation function model predictions ($\ell=0$: monopole, $\ell=2$: quadrupole, $\ell=4$: hexadecapole) with the mean of CMASS mocks AMF correlation function measurements. The black solid line corresponds to the full model (Eq. \ref{eq:fullmodel}), the green short-dashed line to the \citet{tamone20} model, and the red long-dashed line to the original \citet{burden17} ansatz.}
   \label{fig:modelsLRG}
\end{figure}
 
We present in Fig. \ref{fig:modelsLRG} the comparison between the original ansatz from \citet{burden17}, the model used in \citet{tamone20}, the full model presented in this paper, and mock predictions. We can see that the original ansatz allows recovering only qualitatively the mock AMF correlation function mutipole moments, with a significant shift in amplitude. Conversely, \citet{tamone20} and the full model provide similar predictions, very close to the mock measurement. By looking closely at the differences between these two models, we see that the full model performs best, particularly on the smallest scales of the monopole and on the hexadecapole. It is worth recalling that \citet{tamone20} model has a free parameter, $\bar{\chi}$, which we optimised here to best reproduce the measured mocks AMF correlation function.

\begin{figure} 
     \centering
     \includegraphics[width=\columnwidth]{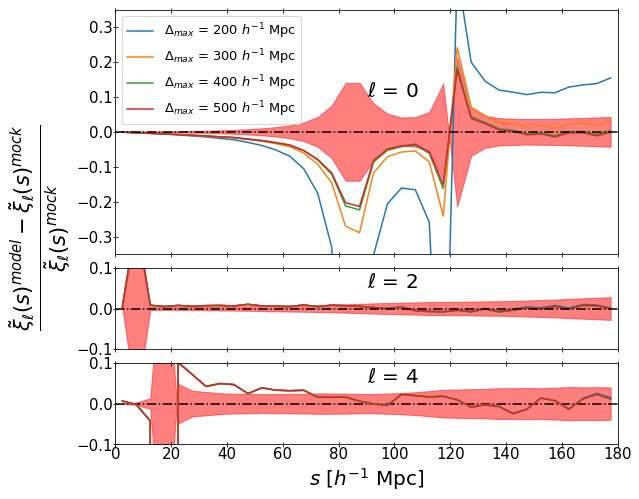}
     \caption{Relative difference between the full model (black solid line in Fig. \ref{fig:modelsLRG}) and the mean of CMASS mocks measurements as a function of $\Delta_{\rm max}$ for the monopole (top), quadrupole (middle), and hexadecapole (bottom) AMF correlation function. The red shade area represents the 1$\sigma$ statistical uncertainty.}
   \label{fig:pimax}  
\end{figure}

In evaluating the models, we have in practice to define the limit of integration for the integral over $s_\parallel$ or $\Delta$ in Eqs. \ref{eq:modelflatsky} and \ref{eq:fullmodel}. The impact of this choice on the full model accuracy is presented in Fig. \ref{fig:pimax}. The latter shows the relative difference of the full model prediction with respect to the mocks prediction, for different values of $\Delta_{\rm max}$ varying from $200$ to $500$ Mpc/h. $\Delta_{\rm max} = 500$ Mpc/h corresponds approximately to the maximum scale possibly probed in the mock survey volume. The red area in Fig. \ref{fig:pimax} represents the 1$\sigma$ deviation around the mean of the mocks. We find that, as expected, by increasing $\Delta_{\rm max}$ the prediction converges to the expected signal, particularly in the monopole. For the quadrupole and hexadecapole, the prediction already converges for $\Delta_{\rm max}=200$ Mpc/h. We note that in this figure, the two strong departures from zero around the BAO peak in the monopole and on small scales in the quadrupole and hexadecapole, are artifacts due to the zero crossing of these functions. Overall, we find in the case of the CMASS sample that $\Delta_{\rm max}=400$ Mpc/h allows the recovery of the mocks prediction within the percent. While the quadrupole signal is retrieved at all scales below 1 $\sigma$, we can see slightly larger shifts in the monopole and hexadecapole. The impact of these shifts on the determination of cosmological parameters are presented in Section \ref{sec:cosmo_analysis}.

\begin{figure} 
     \centering
     \includegraphics[width=\columnwidth]{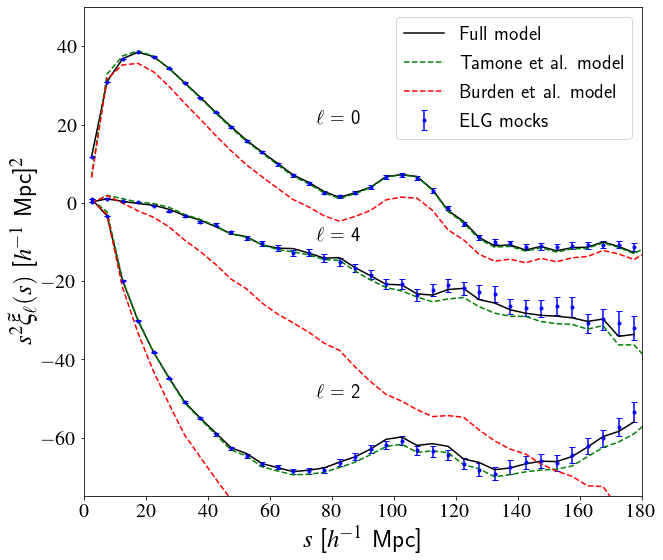}
     \caption{Same as Fig. \ref{fig:modelsLRG} but for ELG mocks.}
   \label{fig:modelsELG}  
 \end{figure}

We repeat the model comparison for the ELG sample, a population of galaxies that has a different intrinsic clustering properties with respect to CMASS. In that case, the correlation function measurements are less precise, mainly due to the smaller volume probed by the ELG mocks. The mean AMF correlation function in the ELG mocks is shown in Fig. \ref{fig:modelsELG} together with the prediction of the full model. Similarly as for CMASS, the agreement is very good, with the full model prediction falling for most of the scales within 1$\sigma$ measurement errors. We note that the ELG sample selection function is more complex than than of CMASS sample, as the sample is made of different sky patches \citep[][]{tamone20} with slightly different $\bar{n}(z)$. In the modelling we use a single averaged $\bar{n}(z)$, which can explain the small differences seen on large scales with respect to the mock prediction and that were not present in the CMASS case. 

\section{Cosmological analysis of CMASS and ELG AMF clustering} \label{sec:cosmo_analysis}

In this section, we investigate the accuracy of the AMF correlation function model in recovering the fiducial cosmological parameters of the mocks, as well as how this compares to the analysis of the standard correlation function multipoles. We perform both a full-shape RSD and BAO-only analysis, similarly as performed in \citet{bautista21} on the eBOSS luminous red galaxy sample.

\subsection{Redshift-space distortions modelling} \label{sec:tns}

The redshift-space correlation function model considered in this work is the \citet{taruya10} model extended to non-linearly biased tracers, hereafter referred to as the TNS model. Its main features are described here and we refer the reader to \citet{bautista21} for a more detailed description.

In this model, the expression for the redshift-space power spectrum of biased tracer is 
\begin{multline}
P^s(k,\nu) = D(k\nu\sigma_v) \big[ P_{\rm gg}(k) + 2\nu^2fP_{\rm{g} \theta}(k) + \nu^4f^2 P_{\theta\theta}(k) + \\
C_A(k,\nu,f,b_1) + C_B(k,\nu,f,b_1) \big],
\label{eq:psg}
\end{multline}
where $k$ is the norm of the wave-vector, $\nu$ is the cosine angle between the wave-vector and the line of sight, $\theta$ is the divergence of the velocity field $\mathbf{v}$ defined as 
$\theta = -\nabla {\bf \cdot v}/(aHf)$, with $f$ the linear growth rate parameter. $P_{gg}$,  $P_{\theta\theta}$ and $P_{g\theta}$ are the galaxy-galaxy, velocity divergence-velocity divergence, and galaxy-velocity divergence power spectra, respectively. The linear matter power spectrum is estimated with {\sc CAMB} at the fiducial cosmology, while non-linear prescriptions for the matter and velocity divergence field are derived with {\sc RESPRESSO} \citet{nishimichi17} and \citet{bel19} fitting functions. We adopted the biasing model of \citet{assassi14} to predict $P_{gg}$ and $P_{g\theta}$, which depends on three bias parameters: $b_1$, $b_2$, $b_{\Gamma_{3}}$ (the additional $b_{\mathcal{G}_{2}}$ parameter is fixed to local Lagrangian prediction). $C_A(k,\nu,f)$ and $C_B(k,\nu,f)$ are the two correction terms given in \citet{taruya10}, which reduce to one-dimensional integrals of the linear matter power spectrum.

The phenomenological damping function $D(k\nu\sigma_v)$ not only describes the Finger-of-God effect induced by random motions in virialized systems, but has also a damping effect on the power spectra. We adopted a Lorentzian form, $D(k,\nu,\sigma_v) = (1+k^2\nu^2\sigma_v^2/2)^{-2}$, where $\sigma_v$ represents an effective pairwise velocity dispersion treated as a nuisance parameter in the cosmological inference. In total, this model has five free parameters $p=[f,b_1,b_2,b_{\Gamma_{3}},\sigma_v]$. The normalization of the input matter power spectrum is set to its fiducial value, $\sigma_8=0.8$, and because of the well-known degeneracy between $f$ and $\sigma_8$, the final constraints are provided in terms of $f$ times the fiducial $\sigma_8$ value at the effective redshift of the sample. 

The TNS correlation function multipole moments are eventually obtained by performing the Hankel transform of the model power spectrum multipole moments,
\begin{equation} \label{eq:multipoleTNS}
\xi^\mathrm{TNS}_\ell(s) = i^\ell \frac{2\ell+1}{2} \int \der k \frac{k^2}{2 \pi^2} j_\ell(ks) \int_{-1}^{1} \der \nu\, P^s(k, \nu) L_\ell(\nu),
\end{equation}
where $j_\ell$ denotes the spherical Bessel function of order $\ell$. In practice, the Hankel transform, i.e. the outer integral in the above equation, is performed rapidly using {\sc FFTLog} algorithm \citep{hamilton00}. The model AMF correlation function multipole moments are evaluated using Eq. \ref{eq:multipolefmodel}, where $\xi(s,\mu)$ in Eqs. \ref{eq:Csmufmodel} and \ref{eq:Asmufmodel} is replaced by $\xi^{\rm TNS}(s,\mu)$. The sum over TNS multipoles is limited to even multipole moments up to $\ell=8$, since the other moments vanish \citep{taruya10,delatorre12}.

We parametrise the Alcock-Paczy\'nski (AP) distortions \citep[][]{alcock79} induced by the assumed fiducial cosmology in the measurements via two dilation parameters that scale transverse, $\alpha_{\perp}$, and radial, $\alpha_{\parallel}$, separations. These quantities are related
to the comoving angular diameter distance, $D_M =
(1+z)D_A(z)$, and Hubble distance, $D_H = c/H(z)$, respectively, as
\begin{align}
\alpha_{\perp} &= \frac{D_M(z_{\rm eff})}{D_M^{\rm fid}(z_{\rm eff})},
\label{eq:aperp} \\
\alpha_{\parallel} &=  \frac{D_H(z_{\rm eff})}{D_H^{\rm fid}(z_{\rm eff})},
\label{eq:apara}
\end{align}
where $c$ is the speed of light in the vacuum, and $z_{\rm eff}$ is the effective redshift of the sample. We apply these dilation parameters to the theoretical TNS power spectrum $P^{s}(k,\nu)$ in Eq. \ref{eq:multipoleTNS}, so that $P^{s}(k,\nu) \rightarrow P^{s}(k',\nu')$ where
\begin{eqnarray}
\label{eq:AP1}k' &=& \frac{k}{\alpha_\perp}\left[1+\nu^2\left(\frac{1}{F_{\rm AP}^2}-1\right)\right]^{1/2},   \\
\label{eq:AP2}\nu'\ & = & \frac{\nu}{F_{\rm AP}}\left[ 1+\nu^2\left(\frac{1}{F_{\rm AP}^2}-1\right)\right]^{-1/2},
\end{eqnarray}
and $F_{\rm AP}=\alpha_\parallel/\alpha_\perp$. In this way, we do not have to recompute the kernels at each iteration of the likelihood analysis. We have checked that this implementation gives unbiased cosmological measurements when fitting the modelled AMF correlation function, which should in principle give the exact same result as the one of the standard correlation function, regardless of the correctness of the considered AMF model. 

In full-shape correlation function or power spectrum analyses, the model real-space power spectrum shape is usually kept fixed at the effective redshift of the sample. However, AP distortions will modify the effective amplitude of the power spectrum, $\sigma_8$, and in turn affect the estimated $f\sigma_8$. Therefore, we introduce the rescaled $f\sigma_8$ parameter as in \citet{Hector2021}, defined as 
\begin{equation}
f\sigma_{8{\text{rs}}} = f\sigma_{8\alpha_{\rm iso}},
\end{equation}
where $\alpha_{\text{iso}} = \aper^{2/3}\apar^{1/3}$ and $\sigma_{8\alpha_{\rm iso}}$ refers to $\sigma_{\rm R}$ evaluated at scale $R=8\alpha_{\rm iso}~h^{-1}~$Mpc. This provides more robust $f\sigma_8$ measurements when the dilation parameters are significantly different from unity, i.e when the fiducial cosmology is far from the true underlying cosmology \citep[see][]{bautista21}.

\subsection{BAO modelling}

In the BAO-only analysis, we follow \citet{bautista21} and use the phenomenological BAO power spectrum model:
\begin{multline}
    P(k, \nu)  = \frac{b^2 \left[1+\beta(1-S(k))\nu^2\right]^2}
{(1+ k^2\nu^2\Sigma_s^2/2)} \\  
\times \left[ P_{\rm no \ peak}(k) + P_{\rm peak}(k)
e^{-k^2\Sigma_{\rm nl}^2(\nu)/2}   \right],
\label{eq:pk2d}
\end{multline}
where $b$ is the linear bias, $\beta = f/b$ is the redshift-space distortions parameter. The non-linear broadening of the BAO peak is modelled by multiplying the 'peak-only' power spectrum $P_{\rm
  peak}$ by a Gaussian with variance $\Sigma_{\rm nl}^2(\nu) =
\Sigma_\parallel^2 \nu^2 + \Sigma^2_\perp(1-\nu^2)$. The nonlinear random motions on small scales are modelled by a Lorentzian distribution parametrized by $\Sigma_s$. The function $S$ models the smoothing of the density field used for the reconstrution. Since we will only consider the pre-reconstrution correlation function here, this is set to zero. The AP dilation parameters are applied only to the peak component of the power spectrum such that 
\begin{multline}
    P(k^\prime, \nu^\prime)  = \frac{b^2 \left[1+\beta(1-S(k))\nu^2\right]^2}
{(1+ k^2\nu^2\Sigma_s^2/2)} \\  
\times \left[ P_{\rm no \ peak}(k) + P_{\rm peak}(k^\prime)
e^{-k^2\Sigma_{\rm nl}^2(\nu)/2}   \right].
\label{eq:pk2d_2}
\end{multline}
The associated correlation function multipole moments are obtained by Hankel transforming the model power spectrum multipole moments as
\begin{equation} \label{eq:multipoleBAO}
\xi_\ell(s) = i^\ell \frac{2\ell+1}{2} \int \der k \frac{k^2}{2 \pi^2} j_\ell(ks) \int_{-1}^{1} \der \nu\, P(k', \nu') L_\ell(\nu).
\end{equation}
The final BAO model is a combination of the BAO correlation function with a smooth function of the separation that allows marginalizing over broadband nonlinear features:
\begin{equation}
\xi^{BAO}_\ell(s) = \xi_\ell(\alpha_{\perp}, \alpha_{\parallel}, s) + 
    \sum_{i=-2}^{0} a_{\ell, i}{s^i}.
\label{eq:template}
\end{equation}
The model for the AMF BAO correlation function multipole moments is obtained from Eq. \ref{eq:multipolefmodel}, where $\xi(s,\mu)$ in Eqs. \ref{eq:Csmufmodel} and \ref{eq:Asmufmodel} is replaced by $\xi^{\rm BAO}(s,\mu)$, and
\begin{equation}
    \xi^\mathrm{BAO}(s,\mu) = \sum_{\ell=0}^{\ell=4} \xi_{\ell}^\mathrm{BAO}(s) L_\ell(\mu).
\end{equation}
In the BAO modelling, a linear RSD model is implicitly assumed, which only predicts non-vanishing even multipole moments up to $\ell = 4$. The implementation of AP dilation parameters is exactly the same as previously for the RSD modelling. In order to model the AMF BAO multipoles, we apply the broadband parameters in Eq. \ref{eq:template} after the AMF modelling. This is mainly to avoid adding two extra broadband parameters to the model hexadecapole, which cannot be well constrained when fitting only the BAO feature in the monopole and quadrupole moments of the correlation function.

\subsection{Full-shape redshift-space distortions results}

\begin{figure} 
     \centering
     \includegraphics[width=\columnwidth]{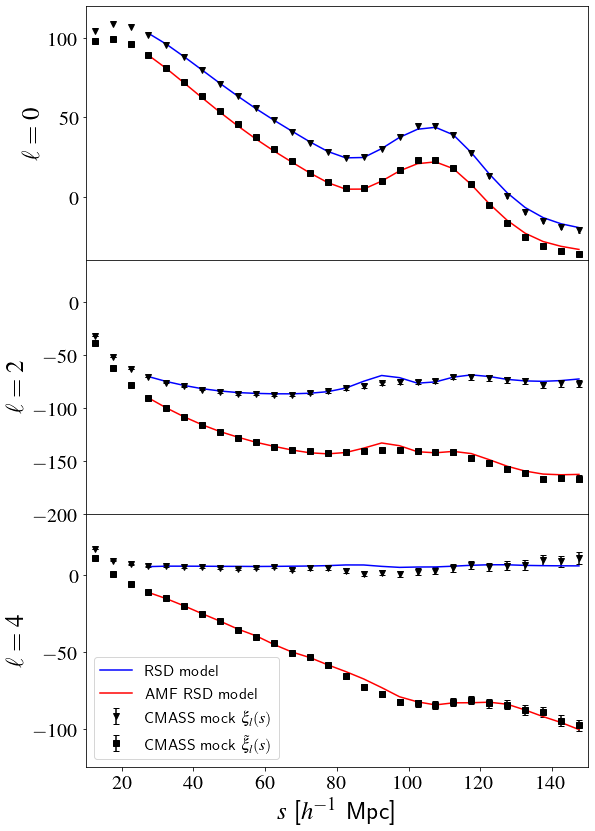}
     \caption{Best-fit RSD models to the mean of the standard and AMF correlation function monopole (top) and quadrupole (middle), and hexadecapole (bottom) measured on the CMASS mocks.}
   \label{fig:bestfitRSD}  
 \end{figure}

We derive constraints on $\apar,\aper, f\sigma_8$ parameters by performing likelihood analyses of both standard and AMF mean correlation functions in the mocks. The monopole, quadrupole, and hexadecapole moments are fitted to the TNS model presented in Section \ref{sec:tns} in the range $25<r<150~h^{-1}$Mpc. AMF kernels are precomputed up to a maximum value of $\Delta_{\rm max} = 500$ Mpc/h, necessary to provide a robust modelling of the BAO feature, as discussed in the previous section. 

%The TNS RSD model breaks down on scales below $20-30$ Mpc/h, depending on the galaxy population and redshift \citep{bautista21,tamone20}. This is common to pertubation theory-based RSD models such as TNS and has an impact on the prediction of the AMF signal. This is because the model anisotropic correlation function needs to be integrated from $s=0$. Therefore for those RSD models, the absolute value of the correction terms can be underestimated. Since our aim is to test the AMF model, independently of the underlying RSD model used, we fix the small-scale amplitude of the multipole moments in the model to the actual mock values (TBD). Nonetheless, in real data analysis we cannot do that and this can be an issue. To tackle this issue \citet{tamone20} apply the AMF model transformation to the data measurements as well. Other possibilities involve supplementing the TNS model with a specific model for the non-linear clustering, for instance a halo model-based prediction \citep{reid14}, or extrapolating small-scale model correlation function amplitude.   

\begin{figure} 
     \centering
     \includegraphics[width=\columnwidth]{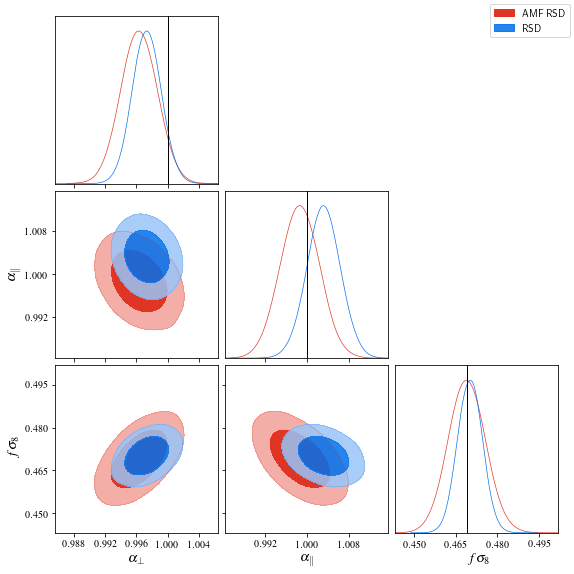}
     \caption{Posterior probability contours on $\aper$, $\aper$ and $f\sigma_8$ obtained when fitting the mean of the standard (blue) and AMF (red) multipole moments in the full-shape RSD analysis of CMASS mocks. The vertical lines in the top panels show the fiducial values of the mocks.}
   \label{fig:contours_RSD}  
\end{figure}

The constraints on $\apar,\aper, f\sigma_8$ that we obtain are given in Table \ref{tab:stats} and the best-fitting models are shown in Fig. \ref{fig:bestfitRSD}. We find that the AMF correlation function, when compared to the standard correlation function, provides unbiased estimate of $f\sigma_8$ and $\aper$. The fiducial values lie within their 1$\sigma$ statistical uncertainty. The recovered $\apar$ central value exhibits a small $1.4\sigma$ shift with respect to the standard analysis, nonetheless it is comparably close to the fiducial value as is the standard analysis. The $1\sigma$ statistical errors obtained on the parameters in the AMF analysis are increased by $20\%$, $10\%$ and $50\%$ for $\apar$, $\aper$, and $f\sigma_8$ respectively. As expected, the signal is decreased in the AMF correlation function leading to worse constraints, the most affected parameter being $f\sigma_8$. The posterior probability contours for all combinations of parameters are shown in Fig. \ref{fig:contours_RSD}. These contours are obtained with the ensemble sampler ZEUS \citep{karamanis2020ensemble,karamanis2021zeus}. We can see that, while AMF analysis shows larger contours compared to the standard one, the directions of degeneracy between the parameters is the same.

For the ELG mocks, we only consider the monopole and quadrupole in the likelihood analysis. Indeed, we found that including the hexadecapole in the standard analysis introduces a 3$\sigma$ shift on $\apar$. Since this shift is not present in the N-Body CMASS mocks, we conclude that we cannot safely compare standard and AMF cosmological measurements when the hexadecapole is included. This is likely due to the approximated method used to produce EZmocks. It is important to emphasise that EZmocks were not meant to reproduce the observed ELG clustering with highest accuracy, instead to reach an accuracy comparable to the statistical precision of the eBOSS ELG sample, where statistical 1$\sigma$ relative precisions on $\apar$, $\aper$, and $f\sigma_8$ are $9.6\%$, $14.7\%$, and $26.3\%$ respectively \citep{ross20,tamone20,raichoor21,demattia21}. Consequently, we can hardly judge from systematic deviations on the parameters below typically 1$\sigma$ in these mocks. 

\begin{figure} 
     \centering
     \includegraphics[width=\columnwidth]{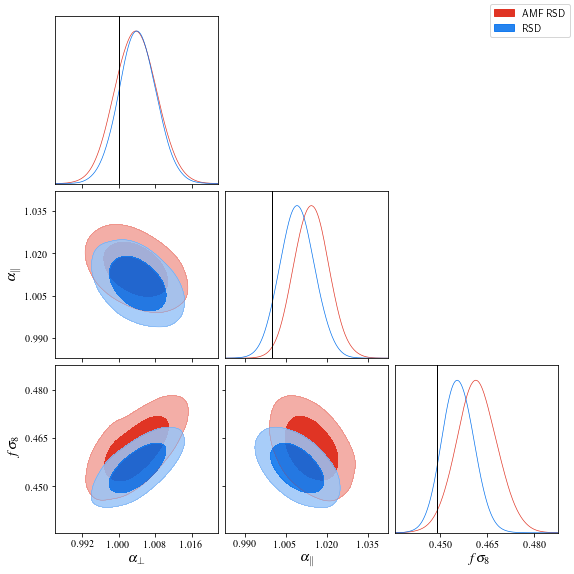}
     \caption{Same as Figure \ref{fig:contours_RSD} but for ELG mocks.}
   \label{fig:Contours_RSD_ELG}
\end{figure}

The ELG constraints on $\apar, \aper, f\sigma_8$ that we obtain are given in Table \ref{tab:stats}. We find that the AMF analysis provides similar constraints on $\aper$ as the standard analysis. The AMF central values for $\apar$ and $f\sigma_8$ lie within 1$\sigma$ of the standard analysis uncertainty. We note that the close to 1$\sigma$ shift in $f\sigma_8$ almost disappear when considering $f\sigma_{8\textrm{rs}}$. The posterior probability contours for all combinations of parameters are shown in Fig. \ref{fig:Contours_RSD_ELG}. As for CMASS mocks, the degeneracy directions between the parameters are similar in the standard and AMF analyses.

\begin{figure} 
     \centering
     \includegraphics[width=\columnwidth]{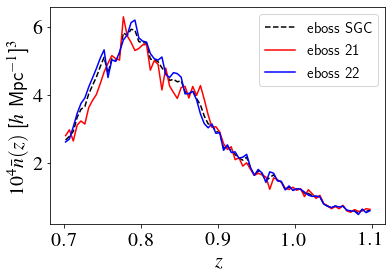}
     \caption{The number density of ELG in the SGC field. The blue (red) curve corresponds to the eboss21 (eboss22) chunk, while the black dashed curve to the average in the SGC field.}
   \label{fig:nz_ELG}  
\end{figure}

\subsection{BAO-only results}

\begin{figure} 
     \centering
     \includegraphics[width=\columnwidth]{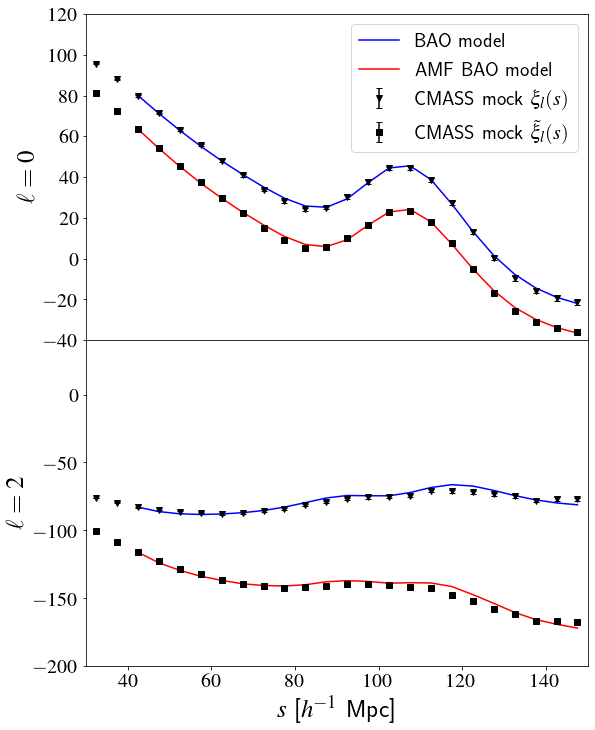}
     \caption{Best-fit BAO models to the mean of the standard and AMF correlation function monopole and quadrupole measured in the CMASS mocks.}
   \label{fig:bestfitBAO}  
\end{figure}

\begin{figure} 
     \centering
     \includegraphics[width=\columnwidth]{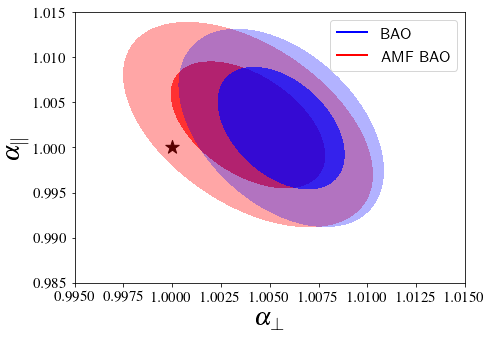}
     \caption{Posterior probability contours on $\aper$, $\aper$ when fitting standard (blue) and AMF (red) multipole moments (monopole and quadrupole only) in the BAO-only analysis on the CMASS mocks.}
   \label{fig:contours_BAO}  
\end{figure}

\begin{figure} 
     \centering
     \includegraphics[width=\columnwidth]{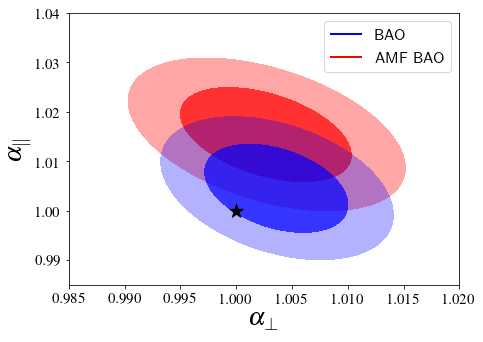}
     \caption{Posterior probability contours on $\aper$, $\aper$ when fitting standard (blue) and AMF (red) multipole moments (monopole and quadrupole only) in the BAO-only analysis on the ELG mocks.}
   \label{fig:contours_BAO_ELG}  
\end{figure}

We further perform a BAO-only analysis on the mean mock AMF and standard correlation functions. We only consider pre-reconstruction correlation functions here, i.e. we do not apply any BAO reconstruction scheme as usually done on real data. Only the monopole and quadrupole are used, as the hexadecapole does not add much constraints in BAO-only analysis \citep{bautista21}. We fit the monopole and quadrupole in the range $40<s<150~h^{-1}$Mpc. The constraints that we obtain on $\apar$ and $\aper$ are given in Table \ref{tab:stats} for the CMASS and ELG samples and the best-fitting CMASS models are shown in Fig. \ref{fig:bestfitBAO}. We find that we can recover almost the same constraints in the AMF and standard analyses in the CMASS mocks. Central values on $\apar$ and $\aper$ are within less than $1\sigma$ of that of the standard analysis. The statistical uncertainty on the parameters increases by less than $5\%$. Overall, the BAO-only analysis of the AMF correlation function is almost as efficient as the standard BAO pre-reconstruction analysis. The joint posterior probability contours for $\apar$ and $\aper$ are shown in Figs. \ref{fig:contours_BAO} and \ref{fig:contours_BAO_ELG} for the CMASS and ELG samples respectively. These contours were computed using the minimisation algorithm {\sc iMinuit}\footnote{\url{https://iminuit.readthedocs.io/}}. 

In the ELG mocks, while the AMF $\aper$ is very close to that obtained in the standard analysis, we observe a $1.6\sigma$ shift on $\apar$. This shift is partially related to the observed shift on the model AMF correlation function shown in Fig. \ref{fig:modelsELG}. It can be explained by the way the AMF model accounts for the varying galaxy radial distribution in the ELG SGC sample. Each ELG SGC mock covers $358 \deg^2$ over an effective volume of 0.5 Gpc$^3$. The SGC footprint is composed of two adjacent chunks: eboss21 and eboss22, respectively covering $117$ and $240$ $\deg^2$. These chunks exhibit slightly different radial distributions as illustrated in Fig. \ref{fig:nz_ELG}. If we perform a BAO-only AMF analysis only on the eboss21 chunk, we find that the shift on $\apar$ disappears. This is shown in Table \ref{tab:statsELG}, where the absolute differences between parameters obtained with the standard and AMF analyses are given. Therefore, the shift on $\apar$ can be attributed to the adopted methodology to derived AMF kernels for the ELG. Overall, and given the intrinsic uncertainty on the clustering in the EZmocks, we can conclude that the AMF does not show any significant bias on the recovered cosmological parameters with respect to the standard analysis.

\begin{table}
  \caption{Absolute difference between the derived parameters in the AMF ELG analysis of the BAO and those obtained in the standard analysis. The results are shown for the case of the entire SGC footprint and that of only the chunk eboss21. The provided errors correspond to the variance obtained by summing up in quadrature the parameter variance from the AMF and standard analyses.}
  \centering
  \begin{tabular}{lcc}
\hline
\hline
Field   &  $ \Delta \alpha_{\perp}$ & 
   $\Delta \alpha_{\parallel}$ \\ % &  $\Delta f\sigma_{8}$ \\
%\hline
%ELG SGC  &&& \\
\hline
%RSD & -0.0010 \pm 0.0065 & 0.0053 \pm 0.0941 & 0.006 \pm 0.084 \\
ELG SGC & $-0.0014 \pm 0.0066$ & $0.0108 \pm 0.0091$ \\ %& \\
%\hline
%ELG eboss21 &&&\\
%\hline
%RSD & -0.0023 \pm 0.0124 & -0.0063 \pm 0.0167 & 0.011 \pm 0.017 \\
ELG eboss21 & $0.0051 \pm 0.0143$ & $0.0009 \pm 0.0175$
\\ %&  \\
\hline
  \end{tabular}
  \label{tab:statsELG}
\end{table}

%We conclude that this will not be an issue for future surveys such as DESI \citep{DESI2016} and Euclid \citep{EUCLID2018} that will observe huge numbers of galaxies over much larger cosmological volumes.

\begin{table*}
  \caption{RSD and BAO results for the mean of NSeries (CMASS sample) and EZmocks (ELG sample). We assume in each analysis the corresponding fiducial cosmology of the mocks. We therefore expect the AP distortions parameters $\apar$ and $\aper$ to be equal to 1. For the growth rate, we expect $f\sigma_8 = 0.469$ and $f\sigma_8=0.449$ for CMASS and ELG respectively. For full-shape analysis, we also present the statistical error corresponding to one realisation after the slash.}
  \begin{tabular}{lcccc}
\hline
\hline
Method   &  $ \alpha_{\perp}$ & 
   $\alpha_{\parallel}$ &  
   $f\sigma_{8}$ & $f\sigma_{8\textrm{rs}}$ \\
\hline
CMASS  &&&& \\
\hline
RSD M+Q+H standard  & $0.9972 \pm (0.0019/0.017)$  & $1.0032 \pm (0.0032/0.029)$ & $
0.4700 \pm (0.0044/0.04)$ & $0.4694 \pm 0.0044$\\
RSD M+Q+H AMF  &  $0.9962 \pm 0.0023$  & $0.9987\pm 0.0038$ & $
0.4686 \pm 0.0067$ & $0.4696 \pm 0.0067$\\
%RSD AMF Eq 15 zeff 0.594 &  $0.9974 \pm 0.0019$ & $0.9874 \pm 0.0027$ & $0.4730 \pm 0.0049$ \\
BAO standard  & $1.0056 \pm 0.0022$  & $1.0007   \pm 0.0044$ & \\
BAO AMF & $1.0043 \pm 0.0026$  & $1.0011   \pm 0.0046$ & \\
\hline
ELG &&&& \\
\hline
RSD M+Q standard  & $1.0038 \pm (0.0043/0.096)$ & $1.0089 \pm   (0.0066/0.147) $ & $ 0.4556 \pm  (0.0053/0.118) $ &  $0.4523 \pm 0.0053$\\
RSD M+Q AMF & $1.0028 \pm 0.0049$ &  $1.0142 \pm 0.0067$ &  $0.4616 \pm 0.0065$ & $0.4567 \pm 0.0065$\\
BAO standard  & $1.0023 \pm 0.0043$  & $1.0063   \pm 0.0062$ & \\
BAO AMF & $1.0009 \pm 0.0051$  & $1.0171 \pm 0.0066$ & \\
\hline
  \end{tabular}
  \label{tab:stats}
\end{table*}

\section{Discussion and conclusion}

In this paper, we studied the use of a modified galaxy two-point correlation function for cosmological inference, whose particularity is to suppress angular modes, and in turn, any potential angular observational systematic errors. This statistic, the AMF two-point correlation function, was first introduced by \citet{burden17}. We extended the latter work and derived a full model to describe this statistic, given a model of the standard redshift-space two-point correlation function. We compared the model to mock galaxy samples of luminous red and emission-line galaxies measurements at $0.43<z<1.1$ and found that it outperforms all previous proposed approximate models. Moreover, it uniquely allows reproducing the full shape of the AMF correlation function, when the underlying correlation function is known, without introducing any new nuisance parameter. This makes possible the performance of a full-shape redshift-space distortions analysis with this statistic.

As a proof of concept, we performed a cosmological analysis of the AMF correlation function in CMASS and ELG mocks, in a similar fashion as we would do with real survey data. We found that we can recover nearly unbiased $\apar$, $\aper$, $f\sigma_8$ parameters with respect to the standard approach. There is only an increase of $18-20\%$ on $\aper$ and $\apar$ statistical uncertainty, and of $50\%$ on $f\sigma_8$ for those galaxy populations.

Current and future large spectroscopic surveys such as DESI \citep{DESI2016} or Euclid \citep{EUCLID2018} will probe much larger universal volumes. This will allow reducing considerably the statistical errors on cosmological parameters. For those, it will be crucial to control the level of systematic errors at a extremely low level. This is today a challenge and the work presented here paves the way towards achieving this goal. By construction, the AMF two-point correlation function is less constraining compared to the standard correlation function. Nonetheless, this approach can be advantageously used in the case of inhomogeneous samples, or in the case of surveys with a complex angular selection function that is poorly understood \citep[e.g.][]{tamone20}. A direct possible application is the cosmological analysis of early-stage dataset from current or future large redshift surveys, such as DESI or Euclid. The latter will suffer from low completeness in the first years of observation \citep{burden17} and the AMF correlation function should allow robust cosmological measurements from these early data.

Finally, it is important to emphasize that the AMF approach is complementary to the standard one, in the sense that it can be used as a cross-validation test. Indeed, it permits studying the impact of angular systematic errors in the standard analysis and be used as a benchmark to check whether all angular systematic errors are well accounted for. If one finds that the recovered cosmological parameters are the same with the two approaches, one validates the accuracy of the observational bias correction scheme used in the standard analysis.

\section*{DATA AVAILABILITY}
The correlation functions, covariance matrices and AMF kernels will be made available after acceptance.
\section*{Acknowledgements}

We thank Michel-Andrès Breton for useful discussions on the paper. The project leading to this publication has received funding from Excellence Initiative of Aix-Marseille University - A*MIDEX, a French "Investissements d'Avenir" programme (AMX-19-IET-008 - IPhU and AMX-20-CE-02). RP, SdlT, SE, and EJ acknowledge the support of the OCEVU Labex (ANR-11-LABX-0060), the French National Research Agency (ANR) under contract ANR-16-CE31-0021, and the A*MIDEX project (ANR-11-IDEX-0001-02) funded by the "Investissements d’Avenir" French government program managed by the ANR. GR acknowledges support from the National Research Foundation of Korea (NRF) through Grants No. 2017R1E1A1A01077508 and No.2020R1A2C1005655 funded by the Korean Ministry of Education, Science and Technology (MoEST). 

Funding for the Sloan Digital Sky Survey IV has been provided by the Alfred P. Sloan Foundation, the U.S. Department of Energy Office of Science, and the Participating Institutions. SDSS-IV acknowledges
support and resources from the Center for High-Performance Computing at
the University of Utah. The SDSS web site is www.sdss.org.

SDSS-IV is managed by the Astrophysical Research Consortium for the 
Participating Institutions of the SDSS Collaboration including the 
Brazilian Participation Group, the Carnegie Institution for Science, 
Carnegie Mellon University, the Chilean Participation Group, the French Participation Group, Harvard-Smithsonian Center for Astrophysics, 
Instituto de Astrof\'isica de Canarias, The Johns Hopkins University, Kavli Institute for the Physics and Mathematics of the Universe (IPMU) / 
University of Tokyo, the Korean Participation Group, Lawrence Berkeley National Laboratory, 
Leibniz Institut f\"ur Astrophysik Potsdam (AIP),  
Max-Planck-Institut f\"ur Astronomie (MPIA Heidelberg), 
Max-Planck-Institut f\"ur Astrophysik (MPA Garching), 
Max-Planck-Institut f\"ur Extraterrestrische Physik (MPE), 
National Astronomical Observatories of China, New Mexico State University, 
New York University, University of Notre Dame, 
Observat\'ario Nacional / MCTI, The Ohio State University, 
Pennsylvania State University, Shanghai Astronomical Observatory, 
United Kingdom Participation Group,
Universidad Nacional Aut\'onoma de M\'exico, University of Arizona, 
University of Colorado Boulder, University of Oxford, University of Portsmouth, 
University of Utah, University of Virginia, University of Washington, University of Wisconsin, 
Vanderbilt University, and Yale University.

%%%%%%%%%%%%%%%%%%%%%%%%%%%%%%%%%%%%%%%%%%%%%%%%%%



\bibliographystyle{mnras}
\bibliography{refs} % if your bibtex file is called example.bib

\begin{thebibliography}{}
\makeatletter
\relax
\def\mn@urlcharsother{\let\do\@makeother \do\$\do\&\do\#\do\^\do\_\do\%\do\~}
\def\mn@doi{\begingroup\mn@urlcharsother \@ifnextchar [ {\mn@doi@}
  {\mn@doi@[]}}
\def\mn@doi@[#1]#2{\def\@tempa{#1}\ifx\@tempa\@empty \href
  {http://dx.doi.org/#2} {doi:#2}\else \href {http://dx.doi.org/#2} {#1}\fi
  \endgroup}
\def\mn@eprint#1#2{\mn@eprint@#1:#2::\@nil}
\def\mn@eprint@arXiv#1{\href {http://arxiv.org/abs/#1} {{\tt arXiv:#1}}}
\def\mn@eprint@dblp#1{\href {http://dblp.uni-trier.de/rec/bibtex/#1.xml}
  {dblp:#1}}
\def\mn@eprint@#1:#2:#3:#4\@nil{\def\@tempa {#1}\def\@tempb {#2}\def\@tempc
  {#3}\ifx \@tempc \@empty \let \@tempc \@tempb \let \@tempb \@tempa \fi \ifx
  \@tempb \@empty \def\@tempb {arXiv}\fi \@ifundefined
  {mn@eprint@\@tempb}{\@tempb:\@tempc}{\expandafter \expandafter \csname
  mn@eprint@\@tempb\endcsname \expandafter{\@tempc}}}

\bibitem[\protect\citeauthoryear{{Alam} et~al.,}{{Alam} et~al.}{2017}]{alam17}
{Alam} S.,  et~al., 2017, \mn@doi [\mnras] {10.1093/mnras/stx721}, \href
  {https://ui-adsabs-harvard-edu.insu.bib.cnrs.fr/abs/2017MNRAS.470.2617A}
  {470, 2617}

\bibitem[\protect\citeauthoryear{{Alam} et~al.,}{{Alam} et~al.}{2021}]{alam21}
{Alam} S.,  et~al., 2021, \mn@doi [\prd] {10.1103/PhysRevD.103.083533}, \href
  {https://ui-adsabs-harvard-edu.insu.bib.cnrs.fr/abs/2021PhRvD.103h3533A}
  {103, 083533}

\bibitem[\protect\citeauthoryear{{Alcock} \& {Paczynski}}{{Alcock} \&
  {Paczynski}}{1979}]{alcock79}
{Alcock} C.,  {Paczynski} B.,  1979, \mn@doi [\nat] {10.1038/281358a0}, \href
  {https://ui-adsabs-harvard-edu.insu.bib.cnrs.fr/abs/1979Natur.281..358A}
  {281, 358}

\bibitem[\protect\citeauthoryear{{Amendola} \& al.}{{Amendola} \&
  al.}{2018}]{EUCLID2018}
{Amendola} L.,  al. 2018, \mn@doi [Living Reviews in Relativity]
  {10.1007/s41114-017-0010-3}, \href
  {https://ui.adsabs.harvard.edu/abs/2018LRR....21....2A} {21, 2}

\bibitem[\protect\citeauthoryear{{Assassi}, {Baumann}, {Green}  \&
  {Zaldarriaga}}{{Assassi} et~al.}{2014}]{assassi14}
{Assassi} V.,  {Baumann} D.,  {Green} D.,   {Zaldarriaga} M.,  2014, \mn@doi
  [\jcap] {10.1088/1475-7516/2014/08/056}, \href
  {https://ui.adsabs.harvard.edu/abs/2014JCAP...08..056A} {2014, 056}

\bibitem[\protect\citeauthoryear{{Bautista} et~al.,}{{Bautista}
  et~al.}{2021}]{bautista21}
{Bautista} J.~E.,  et~al., 2021, \mn@doi [\mnras] {10.1093/mnras/staa2800},
  \href {https://ui.adsabs.harvard.edu/abs/2021MNRAS.500..736B} {500, 736}

\bibitem[\protect\citeauthoryear{{Bel}, {Pezzotta}, {Carbone}, {Sefusatti}  \&
  {Guzzo}}{{Bel} et~al.}{2019}]{bel19}
{Bel} J.,  {Pezzotta} A.,  {Carbone} C.,  {Sefusatti} E.,   {Guzzo} L.,  2019,
  \mn@doi [\aap] {10.1051/0004-6361/201834513}, \href
  {https://ui.adsabs.harvard.edu/abs/2019A&A...622A.109B} {622, A109}

\bibitem[\protect\citeauthoryear{{Blake} et~al.,}{{Blake}
  et~al.}{2010}]{blake10}
{Blake} C.,  et~al., 2010, \mn@doi [\mnras] {10.1111/j.1365-2966.2010.16747.x},
  \href {https://ui.adsabs.harvard.edu/abs/2010MNRAS.406..803B} {406, 803}

\bibitem[\protect\citeauthoryear{{Blake} et~al.,}{{Blake}
  et~al.}{2012}]{blake12}
{Blake} C.,  et~al., 2012, \mn@doi [\mnras] {10.1111/j.1365-2966.2012.21473.x},
  \href {https://ui.adsabs.harvard.edu/abs/2012MNRAS.425..405B} {425, 405}

\bibitem[\protect\citeauthoryear{{Breton} \& {de la Torre}}{{Breton} \& {de la
  Torre}}{2021}]{breton21}
{Breton} M.-A.,  {de la Torre} S.,  2021, \mn@doi [\aap]
  {10.1051/0004-6361/202039603}, \href
  {https://ui.adsabs.harvard.edu/abs/2021A&A...646A..40B} {646, A40}

\bibitem[\protect\citeauthoryear{{Burden}, {Padmanabhan}, {Cahn}, {White}  \&
  {Samushia}}{{Burden} et~al.}{2017}]{burden17}
{Burden} A.,  {Padmanabhan} N.,  {Cahn} R.~N.,  {White} M.~J.,   {Samushia} L.,
   2017, \mn@doi [\jcap] {10.1088/1475-7516/2017/03/001}, \href
  {https://ui-adsabs-harvard-edu.insu.bib.cnrs.fr/abs/2017JCAP...03..001B}
  {2017, 001}

\bibitem[\protect\citeauthoryear{{Chon}, {Challinor}, {Prunet}, {Hivon}  \&
  {Szapudi}}{{Chon} et~al.}{2004}]{Chon2004}
{Chon} G.,  {Challinor} A.,  {Prunet} S.,  {Hivon} E.,   {Szapudi} I.,  2004,
  \mn@doi [\mnras] {10.1111/j.1365-2966.2004.07737.x}, \href
  {https://ui.adsabs.harvard.edu/abs/2004MNRAS.350..914C} {350, 914}

\bibitem[\protect\citeauthoryear{{Chuang} \& al.}{{Chuang} \&
  al.}{2015}]{Chuang2015}
{Chuang} C.-H.,  al. 2015, \mn@doi [\mnras] {10.1093/mnras/stu2301}, \href
  {https://ui.adsabs.harvard.edu/abs/2015MNRAS.446.2621C} {446, 2621}

\bibitem[\protect\citeauthoryear{{Cole} et~al.,}{{Cole} et~al.}{2005}]{cole05}
{Cole} S.,  et~al., 2005, \mn@doi [\mnras] {10.1111/j.1365-2966.2005.09318.x},
  \href {https://ui.adsabs.harvard.edu/abs/2005MNRAS.362..505C} {362, 505}

\bibitem[\protect\citeauthoryear{{DESI Collaboration}}{{DESI
  Collaboration}}{2016}]{DESI2016}
{DESI Collaboration} 2016, arXiv e-prints, \href
  {https://ui.adsabs.harvard.edu/abs/2016arXiv161100036D} {p. arXiv:1611.00036}

\bibitem[\protect\citeauthoryear{{Dawson} et~al.,}{{Dawson}
  et~al.}{2013}]{boss}
{Dawson} K.~S.,  et~al., 2013, \mn@doi [\aj] {10.1088/0004-6256/145/1/10},
  \href {https://ui.adsabs.harvard.edu/abs/2013AJ....145...10D} {145, 10}

\bibitem[\protect\citeauthoryear{{Dawson} et~al.,}{{Dawson}
  et~al.}{2016}]{eboss}
{Dawson} K.~S.,  et~al., 2016, \mn@doi [\aj] {10.3847/0004-6256/151/2/44},
  \href
  {https://ui-adsabs-harvard-edu.insu.bib.cnrs.fr/abs/2016AJ....151...44D}
  {151, 44}

\bibitem[\protect\citeauthoryear{{Fisher}, {Davis}, {Strauss}, {Yahil}  \&
  {Huchra}}{{Fisher} et~al.}{1994}]{fisher94}
{Fisher} K.~B.,  {Davis} M.,  {Strauss} M.~A.,  {Yahil} A.,   {Huchra} J.,
  1994, \mn@doi [\mnras] {10.1093/mnras/266.1.50}, \href
  {https://ui.adsabs.harvard.edu/abs/1994MNRAS.266...50F} {266, 50}

\bibitem[\protect\citeauthoryear{{Gil-Mar{\'\i}n} et~al.,}{{Gil-Mar{\'\i}n}
  et~al.}{2020}]{Hector2021}
{Gil-Mar{\'\i}n} H.,  et~al., 2020, \mn@doi [\mnras] {10.1093/mnras/staa2455},
  \href {https://ui.adsabs.harvard.edu/abs/2020MNRAS.498.2492G} {498, 2492}

\bibitem[\protect\citeauthoryear{{G{\'o}rski}, {Hivon}, {Banday}, {Wandelt},
  {Hansen}, {Reinecke}  \& {Bartelmann}}{{G{\'o}rski}
  et~al.}{2005}]{Gorski2005}
{G{\'o}rski} K.~M.,  {Hivon} E.,  {Banday} A.~J.,  {Wandelt} B.~D.,  {Hansen}
  F.~K.,  {Reinecke} M.,   {Bartelmann} M.,  2005, \mn@doi [\apj]
  {10.1086/427976}, \href
  {https://ui.adsabs.harvard.edu/abs/2005ApJ...622..759G} {622, 759}

\bibitem[\protect\citeauthoryear{{Gunn} et~al.,}{{Gunn} et~al.}{2006}]{gunn06}
{Gunn} J.~E.,  et~al., 2006, \mn@doi [\aj] {10.1086/500975}, \href
  {https://ui.adsabs.harvard.edu/abs/2006AJ....131.2332G} {131, 2332}

\bibitem[\protect\citeauthoryear{{Hahn}}{{Hahn}}{2005}]{hahn04}
{Hahn} T.,  2005, \mn@doi [Computer Physics Communications]
  {10.1016/j.cpc.2005.01.010}, \href
  {https://ui.adsabs.harvard.edu/abs/2005CoPhC.168...78H} {168, 78}

\bibitem[\protect\citeauthoryear{{Hamilton}}{{Hamilton}}{2000}]{hamilton00}
{Hamilton} A.~J.~S.,  2000, \mn@doi [\mnras]
  {10.1046/j.1365-8711.2000.03071.x}, \href
  {https://ui.adsabs.harvard.edu/abs/2000MNRAS.312..257H} {312, 257}

\bibitem[\protect\citeauthoryear{Karamanis \& Beutler}{Karamanis \&
  Beutler}{2020}]{karamanis2020ensemble}
Karamanis M.,  Beutler F.,  2020, arXiv preprint arXiv: 2002.06212

\bibitem[\protect\citeauthoryear{Karamanis, Beutler  \& Peacock}{Karamanis
  et~al.}{2021}]{karamanis2021zeus}
Karamanis M.,  Beutler F.,   Peacock J.~A.,  2021, arXiv preprint
  arXiv:2105.03468

\bibitem[\protect\citeauthoryear{{Kitaura}, {Yepes}  \& {Prada}}{{Kitaura}
  et~al.}{2014}]{kitaura14}
{Kitaura} F.~S.,  {Yepes} G.,   {Prada} F.,  2014, \mn@doi [\mnras]
  {10.1093/mnrasl/slt172}, \href
  {https://ui-adsabs-harvard-edu.insu.bib.cnrs.fr/abs/2014MNRAS.439L..21K}
  {439, L21}

\bibitem[\protect\citeauthoryear{{Landy} \& {Szalay}}{{Landy} \&
  {Szalay}}{1993}]{landy93}
{Landy} S.~D.,  {Szalay} A.~S.,  1993, \mn@doi [\apj] {10.1086/172900}, \href
  {https://ui-adsabs-harvard-edu.insu.bib.cnrs.fr/abs/1993ApJ...412...64L}
  {412, 64}

\bibitem[\protect\citeauthoryear{{Le F{\`e}vre} et~al.,}{{Le F{\`e}vre}
  et~al.}{2003}]{lefevre03}
{Le F{\`e}vre} O.,  et~al., 2003, in {Iye} M.,  {Moorwood} A. F.~M.,  eds,
  Society of Photo-Optical Instrumentation Engineers (SPIE) Conference Series
  Vol. 4841, Instrument Design and Performance for Optical/Infrared
  Ground-based Telescopes. pp 1670--1681, \mn@doi{10.1117/12.460959}

\bibitem[\protect\citeauthoryear{{Lewis} et~al.,}{{Lewis}
  et~al.}{2002}]{lewis02}
{Lewis} I.~J.,  et~al., 2002, \mn@doi [\mnras]
  {10.1046/j.1365-8711.2002.05333.x}, \href
  {https://ui.adsabs.harvard.edu/abs/2002MNRAS.333..279L} {333, 279}

\bibitem[\protect\citeauthoryear{{Limber}}{{Limber}}{1953}]{limber53}
{Limber} D.~N.,  1953, \mn@doi [\apj] {10.1086/145672}, \href
  {https://ui-adsabs-harvard-edu.insu.bib.cnrs.fr/abs/1953ApJ...117..134L}
  {117, 134}

\bibitem[\protect\citeauthoryear{{Nishimichi}, {Bernardeau}  \&
  {Taruya}}{{Nishimichi} et~al.}{2017}]{nishimichi17}
{Nishimichi} T.,  {Bernardeau} F.,   {Taruya} A.,  2017, \mn@doi [\prd]
  {10.1103/PhysRevD.96.123515}, \href
  {https://ui.adsabs.harvard.edu/abs/2017PhRvD..96l3515N} {96, 123515}

\bibitem[\protect\citeauthoryear{{Peacock} et~al.,}{{Peacock}
  et~al.}{2001}]{peacock01}
{Peacock} J.~A.,  et~al., 2001, \nat, \href
  {https://ui.adsabs.harvard.edu/abs/2001Natur.410..169P} {410, 169}

\bibitem[\protect\citeauthoryear{{Peebles}}{{Peebles}}{1980}]{peebles80}
{Peebles} P.~J.~E.,  1980, {The large-scale structure of the universe}.
{Princeton University Press}

\bibitem[\protect\citeauthoryear{{Percival} et~al.,}{{Percival}
  et~al.}{2010}]{percival10}
{Percival} W.~J.,  et~al., 2010, \mn@doi [\mnras]
  {10.1111/j.1365-2966.2009.15812.x}, \href
  {https://ui.adsabs.harvard.edu/abs/2010MNRAS.401.2148P} {401, 2148}

\bibitem[\protect\citeauthoryear{{Pezzotta} et~al.,}{{Pezzotta}
  et~al.}{2017}]{pezzotta17}
{Pezzotta} A.,  et~al., 2017, \mn@doi [\aap] {10.1051/0004-6361/201630295},
  \href {https://ui.adsabs.harvard.edu/abs/2017A&A...604A..33P} {604, A33}

\bibitem[\protect\citeauthoryear{{Pinol}, {Cahn}, {Hand}, {Seljak}  \&
  {White}}{{Pinol} et~al.}{2017}]{pinol17}
{Pinol} L.,  {Cahn} R.~N.,  {Hand} N.,  {Seljak} U.,   {White} M.,  2017,
  \mn@doi [\jcap] {10.1088/1475-7516/2017/04/008}, \href
  {https://ui.adsabs.harvard.edu/abs/2017JCAP...04..008P} {2017, 008}

\bibitem[\protect\citeauthoryear{{Raichoor} et~al.,}{{Raichoor}
  et~al.}{2021}]{raichoor21}
{Raichoor} A.,  et~al., 2021, \mn@doi [\mnras] {10.1093/mnras/staa3336}, \href
  {https://ui.adsabs.harvard.edu/abs/2021MNRAS.500.3254R} {500, 3254}

\bibitem[\protect\citeauthoryear{{Reid} et~al.,}{{Reid} et~al.}{2016}]{reid16}
{Reid} B.,  et~al., 2016, \mn@doi [\mnras] {10.1093/mnras/stv2382}, \href
  {https://ui.adsabs.harvard.edu/abs/2016MNRAS.455.1553R} {455, 1553}

\bibitem[\protect\citeauthoryear{{Ross} et~al.,}{{Ross} et~al.}{2012}]{ross12}
{Ross} A.~J.,  et~al., 2012, \mn@doi [\mnras]
  {10.1111/j.1365-2966.2012.21235.x}, \href
  {https://ui.adsabs.harvard.edu/abs/2012MNRAS.424..564R} {424, 564}

\bibitem[\protect\citeauthoryear{{Ross} et~al.,}{{Ross} et~al.}{2020}]{ross20}
{Ross} A.~J.,  et~al., 2020, \mn@doi [\mnras] {10.1093/mnras/staa2416}, \href
  {https://ui.adsabs.harvard.edu/abs/2020MNRAS.498.2354R} {498, 2354}

\bibitem[\protect\citeauthoryear{{Shafer} \& {Huterer}}{{Shafer} \&
  {Huterer}}{2015}]{shafer15}
{Shafer} D.~L.,  {Huterer} D.,  2015, \mn@doi [\mnras] {10.1093/mnras/stu2640},
  \href
  {https://ui-adsabs-harvard-edu.insu.bib.cnrs.fr/abs/2015MNRAS.447.2961S}
  {447, 2961}

\bibitem[\protect\citeauthoryear{{Simon}}{{Simon}}{2007}]{simon07}
{Simon} P.,  2007, \mn@doi [\aap] {10.1051/0004-6361:20066352}, \href
  {https://ui-adsabs-harvard-edu.insu.bib.cnrs.fr/abs/2007A&A...473..711S}
  {473, 711}

\bibitem[\protect\citeauthoryear{{Smee} et~al.,}{{Smee}
  et~al.}{2013}]{Smee2013}
{Smee} S.~A.,  et~al., 2013, \mn@doi [\aj] {10.1088/0004-6256/146/2/32}, \href
  {https://ui.adsabs.harvard.edu/abs/2013AJ....146...32S} {146, 32}

\bibitem[\protect\citeauthoryear{{Szapudi}, {Prunet}  \& {Colombi}}{{Szapudi}
  et~al.}{2001}]{Szapudi2001}
{Szapudi} I.,  {Prunet} S.,   {Colombi} S.,  2001, arXiv e-prints, \href
  {https://ui.adsabs.harvard.edu/abs/2001astro.ph..7383S} {pp
  astro--ph/0107383}

\bibitem[\protect\citeauthoryear{{Tamone} et~al.,}{{Tamone}
  et~al.}{2020}]{tamone20}
{Tamone} A.,  et~al., 2020, \mn@doi [\mnras] {10.1093/mnras/staa3050}, \href
  {https://ui.adsabs.harvard.edu/abs/2020MNRAS.499.5527T} {499, 5527}

\bibitem[\protect\citeauthoryear{{Taruya}, {Nishimichi}  \& {Saito}}{{Taruya}
  et~al.}{2010}]{taruya10}
{Taruya} A.,  {Nishimichi} T.,   {Saito} S.,  2010, \mn@doi [\prd]
  {10.1103/PhysRevD.82.063522}, \href
  {https://ui.adsabs.harvard.edu/abs/2010PhRvD..82f3522T} {82, 063522}

\bibitem[\protect\citeauthoryear{{Tegmark} et~al.,}{{Tegmark}
  et~al.}{2006}]{tegmark06}
{Tegmark} M.,  et~al., 2006, \mn@doi [\prd] {10.1103/PhysRevD.74.123507}, \href
  {https://ui.adsabs.harvard.edu/abs/2006PhRvD..74l3507T} {74, 123507}

\bibitem[\protect\citeauthoryear{{Zhao} et~al.,}{{Zhao} et~al.}{2020}]{zhao20}
{Zhao} C.,  et~al., 2020, arXiv e-prints, \href
  {https://ui-adsabs-harvard-edu.insu.bib.cnrs.fr/abs/2020arXiv200708997Z} {p.
  arXiv:2007.08997}

\bibitem[\protect\citeauthoryear{{de Mattia} \& {Ruhlmann-Kleider}}{{de Mattia}
  \& {Ruhlmann-Kleider}}{2019}]{demattia19}
{de Mattia} A.,  {Ruhlmann-Kleider} V.,  2019, \mn@doi [\jcap]
  {10.1088/1475-7516/2019/08/036}, \href
  {https://ui-adsabs-harvard-edu.insu.bib.cnrs.fr/abs/2019JCAP...08..036D}
  {2019, 036}

\bibitem[\protect\citeauthoryear{{de Mattia} et~al.,}{{de Mattia}
  et~al.}{2021}]{demattia21}
{de Mattia} A.,  et~al., 2021, \mn@doi [\mnras] {10.1093/mnras/staa3891}, \href
  {https://ui.adsabs.harvard.edu/abs/2021MNRAS.501.5616D} {501, 5616}

\bibitem[\protect\citeauthoryear{{de la Torre} \& {Guzzo}}{{de la Torre} \&
  {Guzzo}}{2012}]{delatorre12}
{de la Torre} S.,  {Guzzo} L.,  2012, \mn@doi [\mnras]
  {10.1111/j.1365-2966.2012.21824.x}, \href
  {https://ui.adsabs.harvard.edu/abs/2012MNRAS.427..327D} {427, 327}

\bibitem[\protect\citeauthoryear{{de la Torre} et~al.,}{{de la Torre}
  et~al.}{2013}]{delatorre13}
{de la Torre} S.,  et~al., 2013, \mn@doi [\aap] {10.1051/0004-6361/201321463},
  \href {https://ui.adsabs.harvard.edu/abs/2013A&A...557A..54D} {557, A54}

\makeatother
\end{thebibliography}

%%%%%%%%%%%%%%%%%%%%%%%%%%%%%%%%%%%%%%%%%%%%%%%%%%

%%%%%%%%%%%%%%%%% APPENDICES %%%%%%%%%%%%%%%%%%%%%
\appendix

\onecolumn

\section{Derivation of the AMF two-point correlation function}
\label{sec:appendixA}

\begin{figure}
    \centering
    \includegraphics[width=0.9\columnwidth]{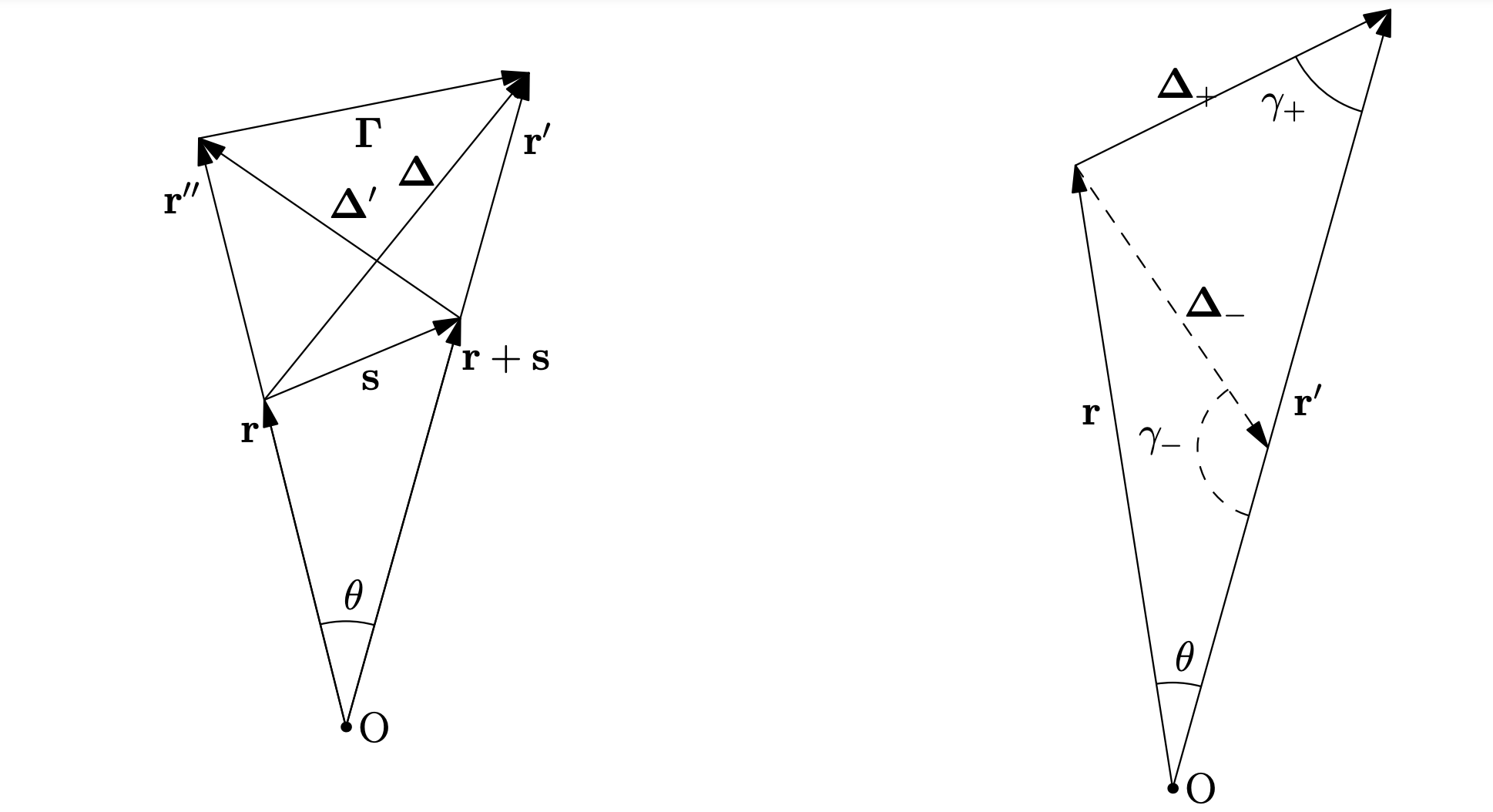}
    \caption{Left: geometrical setting. Right: illustration of the two solutions of the triangle defined by the vectors $(\mathbf{r}, \mathbf{\Delta}, \mathbf{r'})$. For given $r$, $\theta$, and $\Delta = |\mathbf{\Delta_-}| = |\mathbf{\Delta_+}|$, it exists two possible lengths for $r'$ and associated included angle: $\gamma_+$ or $\gamma_-$.}
    \label{fig:geometry}
\end{figure}

The modified Landy-Szalay estimator in Eq. \ref{eq:LSM} is sensitive to the auto-correlation of the AMF overdensity times the survey window function, divided by the window  correlation function. If we define the windowed AMF overdensity field as
\begin{equation}
    F(\mathbf{r}) = P(\mathbf{r})\delta(\mathbf{r}) - P(\mathbf{r}) \int \der r'' \bar{n}(r'') \delta(\mathbf{r}''), 
\end{equation}
where $\mathbf{r}$ and $\mathbf{r}''$ share the same line of sight, the AMF correlation function is
\begin{equation} \label{eq:axitilde}
    \tilde{\xi}(\mathbf{s}) \equiv \frac{\int \der^3 r \, F(\mathbf{r}) F(\mathbf{r}+\mathbf{s})}{\int \der^3 r \, P(\mathbf{r}) P(\mathbf{r}+\mathbf{s})}.
\end{equation}
From the definition of $F(\mathbf{r})$, we have that
\begin{equation} \label{eq:Fdev}
    \begin{split}
    F(\mathbf{r}) F(\mathbf{r}+\mathbf{s}) &=  P(\mathbf{r}) P(\mathbf{r}+\mathbf{s}) \delta(\mathbf{r}) \delta(\mathbf{r}+\mathbf{s}) \\
    & - P(\mathbf{r}) \delta(\mathbf{r}) P(\mathbf{r}+\mathbf{s}) \int \der r'  \bar{n}(r') \delta(\mathbf{r}') \\
    & - P(\mathbf{r}+\mathbf{s}) \delta(\mathbf{r}+\mathbf{s}) P(\mathbf{r})  \int \der r''  \bar{n}(r'') \delta(\mathbf{r}'') \\
    & + P(\mathbf{r}) P(\mathbf{r}+\mathbf{s})
    \int \der r''  \bar{n}(r'') \delta(\mathbf{r}'')
    \int \der r'  \bar{n}(r') \delta(\mathbf{r}'),
    \end{split}
\end{equation}
where $\mathbf{r}'$ and $\mathbf{r}$ + $\mathbf{s}$ are colinear (as well as  $\mathbf{r}$ and $\mathbf{r''}$). By taking the volume integral of Eq. \ref{eq:Fdev}  we can identify four terms. The first one corresponds to the windowed correlation function, the second and third are associated to the cross term in the following, and the fourth term to the angular term. The denominator of Eq. \ref{eq:axitilde} is the window correlation function. Putting all terms together we find that the AMF correlation function can be written
\begin{equation}
    \tilde{\xi}(\mathbf{s}) = \xi(\mathbf{s}) - \frac{C(\mathbf{s})}{W(\mathbf{s})} +
    \frac{A(\mathbf{s})}{W(\mathbf{s})},
\end{equation}
where
\begin{align}
    \xi(\mathbf{s}) &= \left< \delta(\mathbf{r}) \delta(\mathbf{r}+\mathbf{s}) \right> \\
    C(\mathbf{s}) &= \int \der^3 r \, P(\mathbf{r}) P(\mathbf{r}+\mathbf{s}) \left[ 
    \int \der r'  \bar{n}(r') \xi(\mathbf{r}'-\mathbf{r})
    + \int \der r''  \bar{n}(r'') \xi\left(\mathbf{r}''-\mathbf{r}-\mathbf{s}\right) \right], \\
    A(\mathbf{s}) &= \int \der^3 r \, P(\mathbf{r}) P(\mathbf{r}+\mathbf{s})
    \int \der r''  \bar{n}(r'')
    \int \der r'  \bar{n}(r') \xi(\mathbf{r'}-\mathbf{r''}), \\
    W(\mathbf{s}) &= \int \der^3 r \, P(\mathbf{r}) P(\mathbf{r}+\mathbf{s}).
\end{align}

The geometrical configuration is presented on the left hand side of Figure \ref{fig:geometry}. By introducing $\mathbf{r_s}$, $\mathbf{\Delta}$, $\mathbf{\Delta'}$, and $\mathbf{\Gamma}$ such that
\begin{equation}
    \begin{split}
        \mathbf{r_s} &= \mathbf{r} + \mathbf{s} \\
        \mathbf{\Delta} &= \mathbf{r}' - \mathbf{r} \\
        \mathbf{\Delta}' &= \mathbf{r}'' - \mathbf{r_s} \\
        \mathbf{\Gamma} &= \mathbf{r}' - \mathbf{r}'',
    \end{split}
\end{equation}
$A$ and $C$ simplify to
\begin{align}
C(\mathbf{s}) &= \int \der^3 r P(\mathbf{r}) P(\mathbf{r_s}) \left[ 
    \int \der r'  \bar{n}(r') \xi(\mathbf{\Delta}) 
    + \int \der r''  \bar{n}(r'') \xi(\mathbf{\Delta}') \right] = C_1(\mathbf{s}) + C_2(\mathbf{s}) \label{eq:Cterm} \\
    A(\mathbf{s}) &= \int \der^3 r P(\mathbf{r}) P(\mathbf{r_s})
    \int \der r''  \bar{n}(r'')
    \int \der r'  \bar{n}(r') \xi(\mathbf{\Gamma}). \label{eq:Aterm}
\end{align}

In order to calculate those terms we have to solve for three triangles with sides $(\mathbf{r},\mathbf{\Delta},\mathbf{r'})$,$(\mathbf{r''},\mathbf{\Delta'},\mathbf{r_s})$, and $(\mathbf{r''},\mathbf{\Gamma},\mathbf{r'})$. We need to express $r'$ and $r''$ as a function of $r$, $r_s$, $\theta$, $\Delta$, $\Delta'$, and $\Gamma$. This involves solving the general case when two sides and a non-included angle of a triangle are given. In this configuration, there is one or two solutions depending on the relative size of the two given sides. However, since separations are generally smaller than the distance to galaxies, we are always in the case with two solutions. The two degenerate solutions are illustrated on the right hand side of Figure \ref{fig:geometry}. We thus have
\begin{align}
        r' &= r \cos \theta \pm \sqrt{\Delta^2 - r^2 \sin^2 \theta}, \\
        r'' &= r_s \cos \theta \pm \sqrt{\Delta'^2 - r_s^2 \sin^2 \theta}, \\
        r' &= r'' \cos \theta \pm \sqrt{\Gamma^2 - r''^2 \sin^2 \theta}.
\end{align}
Using those relations in Eq. \ref{eq:Cterm} and \ref{eq:Aterm} leads to
\begin{align}
    C_1(\mathbf{s}) &= \int \der^3 r P(\mathbf{r}) P(\mathbf{r_s})
    \int_{-r}^{\infty} \der\Delta \, \frac{\pm \Delta}{\sqrt{\Delta^2 - r^2 \sin^2 \theta}} \, \bar{n}\left(r \cos \theta \pm \sqrt{\Delta^2 - r^2 \sin^2 \theta}\right) \, \xi(\mathbf{\Delta}) \label{eq:c1s}, \\
    C_2(\mathbf{s}) &= \int \der^3 r P(\mathbf{r}) P(\mathbf{r_s})
    \int_{-r_s}^{\infty}  \der\Delta' \, \frac{\pm \Delta'}{\sqrt{\Delta'^2 - r_s^2 \sin^2 \theta}} \,  \bar{n}\left(r_s \cos \theta \pm \sqrt{\Delta'^2 - r_s^2 \sin^2 \theta}\right) \, \xi(\mathbf{\Delta'}), \\
    A(\mathbf{s}) &= \int \der^3 r P(\mathbf{r}) P(\mathbf{r_s})
    \int_0^{\infty} \der r''  \, \bar{n}(r'')
    \int_{-r''}^{\infty}  d\Gamma \, \frac{\pm \Gamma}{\sqrt{\Gamma^2 - r''^2 \sin^2{\theta}}} \, \bar{n}\left(r'' \cos{\theta} \pm \sqrt{\Gamma^2 - r''^2\sin^2{\theta}}\right) \, \xi(\mathbf{\Gamma}).
\end{align}
If we expand $\xi(\mathbf{\Delta})$ in multipole moments in the local plane-parallel approximation such that 
$\xi(\mathbf{\Delta}) = \sum_{p=0}^{\infty} \xi_p(\Delta) L_\ell(\mu_\Delta)$, the term $C_1$ can be re-expressed as
\begin{equation}
    C_1(\mathbf{s}) = 2 \int_0^{\infty} \der\Delta \sum_{p=0}^{\infty} \Delta \, \xi_p(\Delta)
    \int \der^3 r P(\mathbf{r}) P(\mathbf{r_s})
    \frac{\pm \bar{n}\left(r \cos \theta \pm \sqrt{\Delta^2 - r^2 \sin^2 \theta}\right)}{\sqrt{\Delta^2 - r^2 \sin^2{\theta}}} L_p(\mu_{\Delta\pm}).
\end{equation}
By making the change of variable $\Delta \rightarrow -\Delta$ in Eq. \ref{eq:c1s}, the solution with smallest $r'$ (with minus sign) becomes
\begin{align}
    C_{1-}(\mathbf{s}) &= 2 \int_0^{\infty} \der\Delta \sum_{p=0}^{\infty} \Delta \, \xi_p(\Delta)
    \int \der^3 r P(\mathbf{r}) P(\mathbf{r_s})
    \frac{\bar{n}\left(r \cos \theta - \sqrt{\Delta^2 - r^2 \sin^2 \theta}\right)}{\sqrt{\Delta^2 - r^2 \sin^2{\theta}}} L_p(\mu_{\Delta-}).
\end{align}
Therefore, assuming that the two geometrical configurations are equiprobable, we can take the average contribution and
\begin{equation}
    C_1(\mathbf{s}) = \int_0^{\infty} \der\Delta \sum_{p=0}^{\infty} \Delta \, \xi_p(\Delta)
    \int \der^3 r P(\mathbf{r}) P(\mathbf{r_s})
    \frac{\bar{n}\left(r \cos \theta + \sqrt{\Delta^2 - r^2 \sin^2 \theta}\right) L_p(\mu_{\Delta+}) +\bar{n}\left(r \cos \theta - \sqrt{\Delta^2 - r^2 \sin^2 \theta}\right) L_p(\mu_{\Delta-}) }{\sqrt{\Delta^2 - r^2 \sin^2{\theta}}}.
\end{equation}
By further expanding in multipole moments, the latter read
\begin{align}
    C_{1\ell}(s) &= \int_0^{\infty} \der\Delta \sum_{p=0}^{\infty} \Delta \, \xi_p(\Delta) \left( \frac{2\ell+1}{2} \int_{-1}^{1} \der \mu_s  
    \int \der^3 r P(\mathbf{r}) P(\mathbf{r_s}) \notag \right. \\
    & \left. \times \frac{\bar{n}\left(r \cos \theta + \sqrt{\Delta^2 - r^2 \sin^2 \theta}\right) L_p(\mu_{\Delta+}) +\bar{n}\left(r \cos \theta - \sqrt{\Delta^2 - r^2 \sin^2 \theta}\right)L_p(\mu_{\Delta-})}{\sqrt{\Delta^2 - r^2 \sin^2{\theta}}}  L_\ell(\mu_s) \right).
\end{align}
Carrying out in a similar fashion with the other terms we eventually find that
\begin{align}
    X_{\ell}(s) &= \int_0^{\infty} \der \Delta \sum_{p=0}^{\infty} \Delta \, \xi_p(\Delta) \mathcal{W}_{X \ell p}(s,\Delta),
\end{align}
where $X$ stands for $C_1$, $C_2$, or $A$ and
\begin{align}
    \mathcal{W}_{C_1 \ell p}(s,\Delta) &= \frac{2\ell+1}{2} \int_{-1}^{1} \der \mu_s  
    \int \der^3 r \, P(\mathbf{r}) P(\mathbf{r_s}) \notag \\
    & \times \frac{\bar{n}\left(r \cos \theta + \sqrt{\Delta^2 - r^2 \sin^2 \theta}\right) L_p(\mu_{\Delta-})+\bar{n}\left(r \cos \theta - \sqrt{\Delta^2 - r^2 \sin^2 \theta}\right)L_p(\mu_{\Delta-})}{\sqrt{\Delta^2 - r^2 \sin^2{\theta}}}  L_\ell(\mu_s) \label{eq:kernelc1}, \\
    \mathcal{W}_{C_2 \ell p}(s,\Delta) &= \frac{2\ell+1}{2} \int_{-1}^{1} \der \mu_s  
    \int \der^3 r \, P(\mathbf{r}) P(\mathbf{r_s}) \notag \\
    & \times \frac{\bar{n}\left(r_s \cos \theta + \sqrt{\Delta^2 - r_s^2 \sin^2 \theta}\right)L_p(\mu_{\Delta'+}) +\bar{n}\left(r_s \cos \theta - \sqrt{\Delta^2 - r_s^2 \sin^2 \theta}\right)L_p(\mu_{\Delta'-})}{\sqrt{\Delta^2 - r_s^2 \sin^2{\theta}}} L_\ell(\mu_s) \label{eq:kernelc2}, \\
    \mathcal{W}_{A \ell p}(s,\Delta) &= \frac{2\ell+1}{2} \int_{-1}^{1} \der \mu_s  
    \int \der^3 r \, P(\mathbf{r}) P(\mathbf{r_s})
    \int_0^\infty \der r' \bar{n}(r') \notag  \\
    & \times \frac{\bar{n}\left(r' \cos \theta + \sqrt{\Delta^2 - r'^2 \sin^2 \theta}\right)L_p(\mu_{\Gamma+})+\bar{n}\left(r' \cos \theta - \sqrt{\Delta^2 - r'^2 \sin^2 \theta}\right)L_p(\mu_{\Gamma-})}{\sqrt{\Delta^2 - r'^2 \sin^2{\theta}}} L_\ell(\mu_s). \label{eq:kernela}
\end{align}
Similarly, the window correlation function multipole moments can be written as
\begin{equation} \label{eq:kernelw}
    \mathcal{W}_\ell(s) = \frac{2\ell+1}{2} \int_{-1}^{1} \der \mu_s \int \der^3 r P(\mathbf{r}) P(\mathbf{r_s}) L_\ell(\mu_s).
\end{equation}
In the above multipole expansions, we use the mid-point line-of-sight convention so that $\mu_\Delta$, $\mu_{\Delta'}$, $\mu_\Gamma$ correspond to the cosine angles between the separation vectors and their mid-point directions, e.g. $\mu_s = \hat{\mathbf{p}}_s \cdot \hat{\mathbf{s}}$, where $\mathbf{p}_s$ is the position of the median point of $\mathbf{s}$. Those cosine angles can be deduced by solving the half triangle delineated by the median point direction. We thus have that
\begin{align}
    %\mu_{s\pm} &= \frac{\pm r_s \cos \gamma -s/2}{\sqrt{r_s^2 + s^2/4 \mp r_s s \cos \gamma }} \,\,\, {\rm with} \,\,\,
    %\gamma = \arcsin \frac{r \sin \theta}{s}, \\
    \mu_{\Delta\pm} &= \frac{\pm r' \cos \gamma_\Delta-\Delta/2}{\sqrt{r'^2 + \Delta^2/4 \mp r' \Delta \cos \gamma_\Delta }}  \,\,\, {\rm with} \,\,\,
    \gamma_{\Delta} = \arcsin \frac{r \sin \theta}{\Delta}, \\
    \mu_{\Delta' \pm} &= \frac{\pm r'' \cos \gamma_{\Delta'}-\Delta'/2}{\sqrt{r''^2 + \Delta'^2/4 \mp r'' \Delta' \cos \gamma_{\Delta'} }}  \,\,\, {\rm with} \,\,\,
    \gamma_{\Delta'} = \arcsin \frac{r_s \sin \theta}{\Delta'}, \\
    \mu_{\Gamma \pm} &= \frac{\pm r' \cos \gamma_\Gamma - \Gamma/2}{ \sqrt{r'^2 + \Gamma^2/4 \mp r' \Gamma \cos \gamma_\Gamma }} \,\,\, {\rm with} \,\,\,
    \gamma_\Gamma = \arcsin \frac{r'' \sin \theta}{\Gamma}.
\end{align}

The kernels in Eqs. \ref{eq:kernelc1}, \ref{eq:kernelc2}, \ref{eq:kernela}, and \ref{eq:kernelw} are purely geometrical functions that depend on the window function and number density of the sample. They are weighted integrals over the observed volume and can be evaluated with a Monte Carlo method by making use of random catalogues as in \citet{demattia19}. The weighting of each pair is given by the functions in the inner integrands in Eqs. \ref{eq:kernelc1}, \ref{eq:kernelc2}, \ref{eq:kernela}, and \ref{eq:kernelw}. For instance in the case of $\mathcal{W}_\ell$, an estimate can be obtained by computing
\begin{equation}
    \mathcal{W}_{\ell}(s) = \frac{2\ell +1}{2} \frac{1}{2 \pi s^2 \Delta s \, N_{\rm p}}\sum_{i,j} \delta_D(|\mathbf{r}_j-\mathbf{r}_i|-s) L_\ell(\mu_s),
\end{equation}
where the sum goes over the $i^{\rm th}$ and $j^{\rm th}$ random objects of the catalogue, $\Delta s$ is the bin size in $s$, $N_p$ is total number of pairs, and $\delta_D$ denotes the Dirac delta function. An alternative and efficient method to calculate those kernels in a realistic survey configuration, i.e. in a lightcone, is to follow the method of \citet{breton21}. This methods uses two nested spherical volume integrals to sample all pair configurations in the lightcone. It necessitates the a priori knowledge of the angular selection correlation function, which can be efficiently estimated with angular maps of the survey, as well as the mean number density of galaxies as function of distance. Applying a similar approach to the case of the above kernels involves computing the following integrals numerically: 
\begin{align}
    \mathcal{W}_{C_1 \ell p}(s_{\rm min},s_{\rm max},\Delta) &= \frac{2\ell+1}{2} \int_0^{\infty} \der r \, r^2 \bar{n}(r) \int_{s_{\rm min}}^{s_{\rm max}} \der s \int_{-1}^{1} \der \mu \, \Phi(\theta) \, \bar{n}(r_s) \notag \\
    & \times \frac{\bar{n}\left(r \cos \theta + \sqrt{\Delta^2 - r^2 \sin^2 \theta}\right) L_p(\mu_{\Delta-})+\bar{n}\left(r \cos \theta - \sqrt{\Delta^2 - r^2 \sin^2 \theta}\right)L_p(\mu_{\Delta-})}{\sqrt{\Delta^2 - r^2 \sin^2{\theta}}}  L_\ell(\mu_s), \\
    \mathcal{W}_{C_2 \ell p}(s_{\rm min},s_{\rm max},\Delta) &=  \frac{2\ell+1}{2} 
    \int_0^{\infty} \der r \, r^2 \bar{n}(r) \int_{s_{\rm min}}^{s_{\rm max}} \der s \int_{-1}^{1} \der \mu \, \Phi(\theta) \, \bar{n}(r_s) \notag \\
    & \times \frac{\bar{n}\left(r_s \cos \theta + \sqrt{\Delta^2 - r_s^2 \sin^2 \theta}\right)L_p(\mu_{\Delta'+}) +\bar{n}\left(r_s \cos \theta - \sqrt{\Delta^2 - r_s^2 \sin^2 \theta}\right)L_p(\mu_{\Delta'-})}{\sqrt{\Delta^2 - r_s^2 \sin^2{\theta}}} L_\ell(\mu_s), \\
    \mathcal{W}_{A \ell p}(s_{\rm min},s_{\rm max},\Delta) &= \frac{2\ell+1}{2}  \int_0^{\infty} \der r \, r^2 \bar{n}(r) \int_{ s_{\rm min}}^{s_{\rm max}} \der s \int_{-1}^{1} \der\mu \, \Phi(\theta) \, \bar{n}(r_s) \int_0^{\infty} \der r' \bar{n}(r') \notag  \\
    & \times \frac{\bar{n}\left(r' \cos \theta + \sqrt{\Delta^2 - r'^2 \sin^2 \theta}\right)L_p(\mu_{\Gamma+})+\bar{n}\left(r' \cos \theta - \sqrt{\Delta^2 - r'^2 \sin^2 \theta}\right)L_p(\mu_{\Gamma-})}{\sqrt{\Delta^2 - r'^2 \sin^2{\theta}}} L_\ell(\mu_s),
\end{align}
where $\Phi(\theta)$ is the angular selection function correlation function, $\theta$ and $r_s$ can be expressed in terms of $r$, $s$, and $\mu_s$, and $(s_{\rm min},s_{\rm max})$ defines the $s$ bin under consideration. These three- and four-dimensional integrals can be computed efficiently using the {\sc CUBA} library \citep{hahn04} as described in \citet{breton21}. Finally, the AMF anisotropic correlation function can be expressed as
\begin{equation}
    \tilde{\xi}(s,\mu) = \xi(s,\mu) - \frac{C(s,\mu)}{W(s,\mu)} +
    \frac{A(s,\mu)}{W(s,\mu)}
\end{equation}
where
\begin{align}
    C(s,\mu) &= \sum_{\ell=0}^{\infty} \left( \int_0^\infty \der \Delta \sum_{p=0}^{\infty} \Delta \, \xi_p(\Delta) \mathcal{W}_{C \ell p}(s,\Delta)\right) L_\ell(\mu) \\
    A(s,\mu) &= \sum_{\ell=0}^{\infty} \left( \int_0^\infty \der \Delta \sum_{p=0}^{\infty} \Delta \, \xi_p(\Delta) \mathcal{W}_{A \ell p}(s,\Delta) \right) L_\ell(\mu) \\
    W(s,\mu) &= \sum_{\ell=0}^{\infty} \mathcal{W}_{\ell}(s) L_\ell(\mu).
\end{align}
and 
\begin{equation}
    \mathcal{W}_{C \ell p}(s,\Delta) = \mathcal{W}_{C_1 \ell p}(s,\Delta) + \mathcal{W}_{C_2 \ell p}(s,\Delta).
\end{equation}
The multipole moments of the AMF correlation function are obtained from $\tilde{\xi}(s,\mu)$ as
\begin{equation}
\tilde{\xi}_\ell(s) = \frac{(2\ell+1)}{2} \int_{-1}^{1} \tilde{\xi}(s,\mu) L_\ell(\mu)\rm{d} \mu.
\end{equation}

% Don't change these lines
\bsp	% typesetting comment
\label{lastpage}
\end{document}